\documentclass[floatfix,a4paper,aps,prd,twocolumn,preprintnumbers]{revtex4} 

\usepackage{epsfig}
\usepackage{amsmath}
\usepackage{amssymb}
\usepackage{amsfonts}
\usepackage{graphicx}
\usepackage{epic}
\usepackage{overpic}
\usepackage{graphics,subfigure}
\usepackage{float}
\usepackage{booktabs}
\usepackage{color}
\usepackage{geometry}
\usepackage{ulem}
\usepackage{rotating}

\newbox\charbox
\newbox\slabox
\def\s#1{{      
    \setbox\charbox=\hbox{$#1$}
    \setbox\slabox=\hbox{$/$}
    \dimen\charbox=\ht\slabox
    \advance\dimen\charbox by -\dp\slabox
    \advance\dimen\charbox by -\ht\charbox
    \advance\dimen\charbox by \dp\charbox
    \divide\dimen\charbox by 2
    \raise-\dimen\charbox\hbox to \wd\charbox{\hss/\hss}
    \llap{$#1$}0
}}

\newcommand{\newc}{\newcommand}
\newc{\ord}{\mathcal{O}}
\newc{\kap}{\kappa}
\newc{\lam}{\lambda}
\newc{\lamp}{\lambda'}
\newc{\lampp}{\lambda''}
\newc{\beq}{\begin{equation}}
\newc{\eeq}{\end{equation}}
\newc{\barr}{\begin{eqnarray}}
\newc{\earr}{\end{eqnarray}}
\newc{\bit}{\begin{itemize}}
\newc{\eit}{\end{itemize}}
\newc{\tm}{\textrm}
\newc{\gsim}{\stackrel{>}{\sim}}

\definecolor{grau}{gray}{.5}     

\newc{\mtext}[1]{{\color{blue}  #1}}
\newc{\htext}[1]{{\color{magenta}  #1}}
\newc{\ztext}[1]{{\color{red}  #1}}
\newc{\stext}[1]{{\color{green}  #1}}
\newc{\greytext}[1]{{\color{grau}  #1}}
\newc{\vtext}[1]{{\bf  #1}}  
\newc{\vmtext}[1]{\mathbf{#1}}  

\newc{\bea}{\begin{eqnarray}}
\newc{\eea}{\end{eqnarray}}

\newc{\lag}{\mathcal{L}}
\newc{\spot}{\mathcal{W}}
\newc{\cM}{{\cal M}}
\newc{\rpv}{{{\not\!{R}}_p}}
\newc{\rp}{R_p}
\newc{\bt}{B_3}
\newc{\ph}{P_6}
\newc{\eps}{\epsilon}

\newc{\ssl}{\tilde{L}}
\newc{\sse}{\tilde{E}}
\newc{\ssq}{\tilde{Q}}
\newc{\ssd}{\tilde{D}}

\newc{\mN}{\mathcal{M}_N}
\newc{\mnueff}{\mathcal{M}_{\nu}^{{\rm eff}}}
\newc{\mCharge}{\mathcal{M}_C}
\newc{\mlepeff}{\mathcal{M}_{\ell}^{{\rm eff}}}
\newc{\pmns}{U_{\rm PMNS}}
\newc{\ckm}{U_{\rm CKM}}

\newc{\azero}{A_0}
\newc{\mo}{M_0}
\newc{\mhalf}{M_{1/2}}
\newc{\tanb}{{\rm tan}\beta}
\newc{\sgnmu}{{\rm sgn}(\mu)}

\newc{\nubb}{0\nu\beta\beta}

\newc{\al}{\alpha}

\newc{\mgut}{M_X}
\newc{\mz}{M_Z}

\newc{\softsusy}{{\texttt{SOFTSUSY}}}
\newc{\softsusyv}{{\tt SOFTSUSYv3.1.5}}
\newc{\minuit}{{\tt MINUIT2}}

\graphicspath{%
{./}%
{figs/}%
}

\begin{document}

\title{Neutrino masses and mixings in the baryon triality constrained minimal supersymmetric standard model}

\author{Herbi~K.~Dreiner}
\email[]{dreiner@th.physik.uni-bonn.de}
\author{Marja~Hanussek}
\email[]{hanussek@th.physik.uni-bonn.de}
\affiliation{Bethe Center of Theoretical Physics and Physikalisches 
Institut,University of Bonn, Bonn, Germany}

\author{Jong-Soo Kim}
\email[]{jongsoo.kim@tu-dortmund.de}
\affiliation{Institut f\"ur Physik, Technische Universit\"at Dortmund, 
Dortmund, Germany}

\author{C.~H.~Kom} 
\email[]{kom@hep.phy.cam.ac.uk}
\affiliation{Cavendish Laboratory, University of Cambridge, UK
  \\ Department of Applied Mathematics and Theoretical Physics,
  University of Cambridge, UK}

\begin{abstract}
We discuss how the experimental neutrino oscillation data can be
realized in the framework of the baryon triality ($B_3$) constrained
supersymmetric Standard Model (cSSM). We show how to obtain
phenomenologically viable solutions, which are compatible with the
recent WMAP observations. We present results for the hierarchical,
inverted and degenerate cases which illustrate the possible size and
structure of the lepton number violating couplings. We work with a
new, as yet unpublished version of \softsusy\, where we implemented
full one--loop neutrino masses. Finally, we shortly discuss some
phenomenological implications at the LHC.
\end{abstract}

\preprint{Bonn-TH-2011-11}
\preprint{CAVENDISH-HEP-2011-09}
\preprint{DO-TH 11/16}

\maketitle

\section{Introduction}\label{sec:neutrinodata}

Experimentally, it is well established that the Standard Model (SM) of
particle physics requires an extension to accommodate the neutrino
oscillation data
\cite{Davis:1968cp,Fukuda:1998ah,Fukuda:1998fd,Ahmad:2002jz,Aharmim:2005gt,Apollonio:2002gd,Cleveland:1998nv}.
The data indicate that at least two neutrinos are massive and that the
neutrino mixing angles are large.  Many mechanisms have been proposed
to explain the neutrino mass pattern.  The simplest is to introduce
small Dirac mass terms.  However, in order to be compatible with
observations, the Yukawa couplings can at most be of $\ord(10^
{-12})$, which appears highly unnatural.  Furthermore, this requires
additional right--handed neutrinos and the corresponding Majorana mass
terms are unconstrained by SM gauge symmetries.

Alternatively, one can allow for the Majorana mass terms, this is the
so--called (type--I) see--saw mechanism
\cite{Minkowski:1977sc,Mohapatra:1979ia,Yanagida,GellMann:1980vs,Mohapatra:1980yp,Jezabek:1998du}.
By setting the arbitrary Majorana mass scale to be large, light
neutrinos with mass of order $\ord({\rm 0.1\,eV})$ can be obtained
even with $\ord(1)$ Yukawa couplings.  There are other see--saw
mechanisms
\cite{Magg:1980ut,Lazarides:1980nt,Schechter:1980gr,Mohapatra:1980yp,Foot:1988aq,Mohapatra:1986aw,Mohapatra:1986bd,Dreiner:1994ra},
which involve different additional particles that determine/control
the see--saw scale. Some models, involving a see--saw mechanism,
determine the detailed neutrino masses from a broken (gauge) symmetry
\cite{Zee:1980ai,Dreiner:1992zy,Ibanez:1994ig,Dreiner:1994ra,Dreiner:2003yr,He:2003rm,Luhn:2007sy,Dreiner:2007vp}.

In this paper we consider a natural mechanism in supersymmetric
extensions of the SM, which does not require any right--handed, gauge
singlet, neutrinos, or a corresponding new mass scale.  We restrict
ourselves to the minimal particle content, consisting of the SM
particles, an additional Higgs SU(2)$_L$ doublet, and their
superpartners, \textit{i.e.} the supersymmetric Standard Model (SSM)
\cite{Nilles:1983ge,Haber:1984rc,Drees:1996ca}. The most general gauge
invariant and renormalizable SSM Lagrangian contains lepton number
violating (LNV) operators which mix the left--handed neutrinos with the
neutralinos.  The neutralino mass provides a see--saw scale, of
$\ord({\rm 100\,GeV})$, for the generation of light Majorana neutrino
masses~\cite{Hall:1983id,Joshipura:1994ib,Nowakowski:1995dx,Grossman:1997is,Grossman:1999hc,Grossman:2000ex,Nardi:1996iy,Davidson:2000ne,Dedes:2006ni,Dreiner:2007uj,Allanach:2007qc}.

In the generic SSM, there exist LNV and baryon number violating
operators. We thus restrict our model to conserve baryon triality
($B_3$)
\cite{Ibanez:1991hv,Ibanez:1991pr,Banks:1991xj,Dreiner:2005rd,Dreiner:2006xw}.
This prohibits all baryon number violating terms while allowing for
lepton number violation, and the proton is stable.  Furthermore, we
work in the \textit{constrained} baryon triality SSM ($B_3$ cSSM) in
order to limit the number of free (lepton number conserving)
parameters at the unification scale. The relevant details of this
model are presented in the next section.  It is well--known that in the
$B_3$ cSSM, only one light neutrino is massive at tree--level
\cite{Hall:1983id,Joshipura:1994ib,Nowakowski:1995dx,Nardi:1996iy,Davidson:2000ne,Dedes:2006ni,Abada:1999ai,Dreiner:2007uj}.
Higher order corrections need to be included to give mass to at least
one more neutrino in order to be consistent with the non--zero values
of the neutrino mass squared differences, $\Delta m_{21}^2$ and
$\Delta m_{31}^2$.  The radiative origin of the second neutrino mass
scale implies that a strong hierarchy of $\ord(100)$ between the
neutrino masses is to be expected, \textit{cf.}
Ref.~\cite{Allanach:2007qc}. However, the data require a neutrino mass
ratio of the heaviest two neutrinos of at most $\ord(5)$.

Thus a mechanism is needed to suppress the tree--level mass scale for
viable models.  In Ref.\,\cite{Allanach:2007qc}, sets of five
parameters [two trilinear LNV couplings together with three mixing
angles that describe the charged lepton Yukawa matrix] defined in a
cSSM were found to reproduce the oscillation data.  The LNV parameters
were chosen such that their contributions to the tree--level neutrino
masses partially cancel against each other.  Another possibility,
first mentioned in Ref.~\cite{deCarlos:1996du}, is that the
tree--level neutrino mass can vanish in a more generic fashion in
certain regions of cSSM parameter space, specified by the trilinear
soft supersymmetry breaking parameter $A_0$.  A detailed explanation
of how this situation arises, including a discussion on loop
contributions in this parameter space was presented in
Ref.\,\cite{Dreiner:2010ye}.

In this paper, we focus especially on these parameter regions, and aim
to reproduce the neutrino oscillation data using a small set of LNV
couplings. Compared with Ref.\,\cite{Allanach:2007qc}, these regions
might be considered more preferable in the sense that they avoid
suppression of tree level neutrino masses through specific
cancellations between LNV parameters.  Our set--up is also different
from Ref.\,\cite{Allanach:2007qc}, in that we specify the LNV
parameters in a basis where the lepton Yukawa couplings are diagonal.
We consider this advantageous, as this allows for a more transparent
understanding and better control of how different LNV parameters
contribute to the neutrino mass matrix.  Here we also improve on the
numerical calculation performed in Ref.\,\cite{Allanach:2007qc} by
including a full one loop calculation for the sneutrino vacuum
expectation values, on top of the one loop corrections to the neutral
fermion masses.  This computation is implemented as an extension to
the mass spectrum calculational tool
\softsusy~\cite{Allanach:2001kg,Allanach:2009bv}.

Our aim is to obtain the correct masses and mixing angles with a small
number of LNV parameters. We furthermore wish to analyze the general
structures that lead to potential solutions, since it is not possible
to systematically list \textit{all} solutions. This work is an
extension of Ref.~\cite{Dreiner:2010ye}, where single coupling bounds
from the cosmological limit on the neutrino mass,
Eq.~(\ref{bound_numass}), were determined. By introducing parameters
coupled to different generations, we attempt to understand how
different trilinear LNV terms interplay with each other to generate
the observed mass pattern.

The generation of neutrino masses through non--zero LNV parameters
directly at the electroweak scale (therefore without the complications
from renormalisation group effects) has been studied in
Refs.\,\cite{Abada:1999ai,Dedes:2006ni}.  Generation of neutrino
masses via bilinear LNV couplings and the corresponding collider
signatures have also been studied.  We refer interested readers to
Refs.\,\cite{Bhattacharyya:1996nj,Dreiner:1997uz,Barbier:2004ez,Hempfling:1995wj,Kaplan:1999ds,Mira:2000gg,Hirsch:2002ys,Hirsch:2000ef,Diaz:2003as,Bartl:2003uq,Hirsch:2003fe,Hirsch:2004he,Hirsch:2005ag,deCampos:2007bn,deCampos:2008av}
and references therein.

The outline of this paper is as follows.  In section~II, we introduce
the $B_3$ cSSM model, and highlight the most relevant ingredients,
including the choice of benchmark scenarios and (low energy)
observables that could constrain the LNV parameters, for our present
study.  In section III we examine in detail sets of LNV parameters
that can reproduce the neutrino oscillation data.  Section IV is
devoted to the numerical fitting procedure used in our analysis.  We
discuss our best fit parameter sets and possible collider
phenomenology in sections V and VI, before concluding in section VII.

\section{Neutrino Masses in the Baryon Triality ($B_3$) \lowercase{c}SSM}

\subsection{Experimental Neutrino Data}\label{subsec:neutrinodata}
The best fit of the combined
global analysis of atmospheric, solar, reactor and accelerator data in
terms of three active oscillating neutrinos is given by
\cite{Schwetz:2011qt,GonzalezGarcia:2010er}, \bea
\sin^2[\theta_{12}]&=&0.31\pm0.02,\label{eq:data1}\\
\sin^2[\theta_{23}]&=&0.51\pm0.06,\label{eq:data2}\\
\sin^2[\theta_{13}]&<&0.03, \label{eq:data3}\\
\Delta m_{21}^2&=&7.59\pm0.2\times10^{-5}\tm{ eV}^2,\label{eq:data4}\\
\Delta m_{31}^2&=& \left\{
\begin{array}{r}
-2.34\pm0.1\times 10^{-3}\tm{ eV}^2\\
2.45\pm0.1\times 10^{-3}\tm{ eV}^2
\end{array} \right. ,\label{eq:data5}
\eea 
where the errors are given at the $1\sigma$ level, and 
\bea 
\Delta m_{ij}^2 &\equiv& m_i^2-m_j^2.  
\eea 
The data indicate large mixing angles $\theta_{12}$ and $\theta_{23}$
and a small or possibly even vanishing angle $\theta_{13}$.  This
implies that at least two neutrinos have non--zero mass.  The (as--yet)
undetermined sign of $\Delta m_{31}^2$ means that two mass orderings
are possible.  They are known as the normal ($\Delta m_{31}^2 > 0$)
and the inverted ($\Delta m_{31}^2 < 0$) hierarchies.

For illustrative purposes, we often use the tri--bi--maximal mixing
(TBM) approximation \cite{Harrison:2002er}, where
\beq 
\sin^2[\theta_{12}]=
\frac{1}{3},\;\; \sin^2[\theta_{23}]= \frac{1}{2}\,\;\;
\sin^2[\theta_{13}]= 0\label{eq:TBM} 
\eeq 
is assumed. Note that these are all within 1$\,\sigma$ of the
best--fit experimental values given in
Eqs.~(\ref{eq:data1})--(\ref{eq:data3}). The first two
quantities differ from their best fit values by 7\% and 2\%
respectively.  In the TBM approximation, the
Pontecorvo--Maki--Nakagawa--Sakata (PMNS) mixing matrix
\cite{Pontecorvo:1957cp, Maki:1962mu, Pontecorvo:1967fh} is explicitly
given by
\beq U_{TBM}\equiv\left(\begin{array}{ccc}
  \phantom{-}\sqrt{\frac{2}{3}} & \phantom{-}\sqrt{\frac{1}{3}} & \;\;\;\;\;0
  \\ -\sqrt{\frac{1}{6}} & \phantom{-}\sqrt{\frac{1}{3}} &
  \phantom{-}\sqrt{\frac{1}{2}} \\ \phantom{-}\sqrt{\frac{1}{6}} &
  -\sqrt{\frac{1}{3}} & \phantom{-} \sqrt{\frac{1}{2}}
\end{array}
\right)\,.\label{UTBM}
\eeq
Since the defining equations in Eq.~(\ref{eq:TBM}) involve squares,
more than one phase convention exists for the resulting mixing matrix.

The observations and measurements from neutrino oscillations determine
the differences of neutrino masses squared, \textit{cf.}
Eqs.~(\ref{eq:data4}), (\ref{eq:data5}). Direct laboratory
measurements restrict the absolute masses of the neutrinos to be below
$\ord(10\,{\rm MeV} - 1\,{\rm eV})$
\cite{GonzalezGarcia:2010er,GonzalezGarcia:2007ib,Barate:1997zg,Assamagan:1995wb,Bonn:2001tw,Lobashev:2001uu,Amsler:2008zzb}.
Limits dependent on the Majorana nature of neutrinos also exist from
non--observation of neutrinoless double beta decay ($\nubb$), which is
of $\ord(0.5\,{\rm eV})$
\cite{Bakalyarov:2003jk,KlapdorKleingrothaus:2000sn,Arnaboldi:2008ds,Aalseth:2002rf}.
Note, there is a claim of evidence for a neutrino mass of $0.39\,
\mathrm{eV}$ in a $\nubb$ experiment
\cite{KlapdorKleingrothaus:2001ke}.

A stringent upper limit can be obtained from cosmological restrictions
on the sum of the neutrino masses, with the exact limit
dependent on details of the analysis.  Typically these
analyses include data from the Wilkinson Microwave Anisotropy Probe
(WMAP) \cite{Spergel:2006hy}, Large Scale Structure \cite{SDSS,2dFGRS}
and Type Ia supernovae \cite{supernovae}.  For our purpose, we use
\begin{equation}
\sum {m_{\nu_i}} \lesssim 0.4 \, \text{eV} \, ,
\label{bound_numass}
\end{equation}
at $99.9\%$ confidence level, obtained from
Refs.\,\cite{Cirelli:2006kt,Goobar:2006xz}.

In our numerical fitting procedure, we make use of three limiting
cases of neutrino mass hierarchies. In the first two cases, we assume
that the lightest neutrino is massless and impose normal and inverted
hierarchy, respectively. In the third case, we consider
almost--degenerate neutrino masses with normal hierarchy mass ordering,
saturating the cosmological limit stated in Eq.~(\ref{bound_numass}).
 
For the normal ($m_1<m_2<m_3$) and inverted ($m_3<m_1<m_2$)
hierarchies, neutrino masses are respectively given by
\bit
\item normal hierarchy (NH):
\bea
m_1&\approx&0 \tm{ eV},\nonumber\\
m_2&=&8.71\times10^{-3}\tm{ eV},\nonumber\\
m_3&=&4.95\times10^{-2}\tm{ eV}, \nonumber \\
m_3/m_2 &\sim& 5.7\,. \label{eq:NH}
\eea
\item inverted hierarchy (IH):
\bea
m_1&=&4.84\times10^{-2}\tm{ eV},\nonumber\\
m_2&=&4.92\times 10^{-2}\tm{ eV},\nonumber\\
m_3&\approx&0 \tm{ eV}, \nonumber\\
m_2/m_1 &\sim& 1\,. \label{eq:IH}
\eea 
\eit
In our fits, we use the masses given in Eqs.(\ref{eq:NH}) and
(\ref{eq:IH}) as central values for the three neutrino masses for the
NH and IH cases, respectively.  For the degenerate case ($m_1
\approx m_2 \approx m_3$), we assume that the sum of the three active
neutrino masses equals $0.4\tm{ eV}$.

\subsection{Baryon Triality ($B_3$) cSSM}
With the field content of the SSM, the most general gauge invariant superpotential at the
renormalizable level can be written as \cite{Sakai:1981pk,Weinberg:1981wj,Weinberg:1979sa}
\bea 
\spot &=& \spot_{\rp} + \spot_{\rpv},
\eea 
where $\spot_{\rp}$($\spot_{\rpv}$) contain terms that conserve
(violate) the discrete symmetries R--parity ($\rp$) as well as proton
hexality ($\ph$).  In a notation that follows
Ref.~\cite{Allanach:2003eb} and \softsusy\, 
\cite{Allanach:2001kg,Allanach:2009bv} closely, they are
\bea
\spot_{\rp} &=& \eps_{ab} \,[(Y_E)_{jk}H_d^aL_j^b\bar{E}_k + (Y_D)_{jk}H_d^aQ_j^b\bar{D}_k \nonumber \\
  && \phantom{\eps_{ab}}+(Y_U)_{jk}Q_j^aH_u^b\bar{U}_k - \mu H_d^aH_u^b ], \\
\spot_{\rpv} &=& \eps_{ab}\, [\frac{1}{2}\lam_{ijk} L_i^aL_j^b\bar{E}_k +
\lamp_{ijk} L_i^aQ_j^b\bar{D}_k \nonumber \\
  && \phantom{\eps_{ab}} + \lampp_{ijk} \bar{U}_i\bar{D}_j\bar{D}_k - \kappa_i L_i^aH_u^b ],
\eea
where $i,j,k\,\in\{1,2,3\}$ are generation indices, $a,b\in\{1,
2\}$ ($\eps_{12}=1$) are indices of the $SU(2)_L$ fundamental
representation, while the corresponding $SU(3)_c$ indices are
suppressed.  To avoid operators that could result in dangerously fast
proton decay
\cite{Weinberg:1979sa,Wilczek:1979hc,Sakai:1981pk,Smirnov:1996bg}, we
impose the discrete symmetry baryon triality ($\bt$)
\cite{Ibanez:1991hv,Ibanez:1991pr,Banks:1991xj,Dreiner:2005rd,Dreiner:2006xw}.
Under this symmetry, baryon number is conserved while we have lepton
number violation (LNV). The superpotential is given by
\bea
\spot_{\bt} &=& \spot_{\rp} + \spot_{\rm LNV},
\eea
where the last term on the right is obtained by setting $\lampp=0$ in
$\spot_{\rpv}$.  We note that $\rp$, $\bt$ and $\ph$ are the only
discrete symmetries which can be written as a remnant of a broken
anomaly free gauge symmetry
\cite{Ibanez:1991hv,Ibanez:1991pr,Dreiner:2005rd,Banks:1991xj}.  In
the rest of this paper, $\bt$ is assumed to be conserved.

The LNV soft supersymmetry (SUSY) breaking interaction Lagrangian is
given by
\bea \label{eq:softpot}
-\lag_{LNV} &=&\eps_{ab} \,[\frac{1}{2}h_{ijk}\ssl^a_i\ssl^b_j\sse_k 
+h'_{ijk}\ssl^a_i\ssq^b_j\ssd_k \nonumber \\ &&  
- \tilde{D}_i\ssl^a_ih^b_u ] + m^2_{L_iH_d}\ssl^{\dagger}_{ia}h_d^{a} + h.c.,
\eea
where tilde denotes a super--partner of the more familiar Standard
Model field.  The complete $\rp$ soft SUSY breaking Lagrangian can be
found in Ref.~\cite{Allanach:2001kg}.

The $\bt$ SSM model has more than 200 free parameters
\cite{Haber:1997if}.  In order to perform concrete numerical studies,
we restrict our discussion to the cSSM framework \cite{Drees:1996ca}.
The cSSM model is specified by the parameter set
\beq \label{eq:mSUGRAinput}
\mo, \quad \mhalf, \quad \azero, \quad {\rm sgn}(\mu), \quad \tanb,
\eeq
denoting the universal scalar mass, the universal gaugino mass, the
universal trilinear scalar coupling, the sign of the bilinear Higgs
mixing parameter $\mu$ and the ratio of Higgs vacuum expectation
values (VEVs) $v_u/v_d$ at the electroweak scale $\mz$.  Except for
$\tan\beta$, all parameters are defined at the unification scale
$\mgut$.

Additionally, we allow for a subset of $\bt$ conserving (but 
$\rp$--violating) parameters
\beq
{\bf \Lambda} \subset \{ \lam_{ijk},\lamp_{ijk} \}
\eeq
that will be specified in later sections. Note that we allow for
trilinear but not bilinear LNV parameters at the unification scale,
because we work in a basis where the bilinear LNV couplings $\kappa_i$
and $\tilde{D}_i$ are both zero at $\mgut$.  This is possible for
universal SUSY breaking \cite{Allanach:2003eb} via a basis
transformation of the lepton and Higgs superfields \cite{Hall:1983id,
Dreiner:2003hw}.  However, at lower energy scales $\kappa_i$ and
$\tilde{D}_i$ are generated via the renormalization group equations (RGEs)
\cite{Nardi:1996iy}.

\subsection{Neutrino and Charged Lepton Masses}

Since lepton number is violated, the lepton doublet superfields $L_i$
carry the same quantum numbers as the down--type $H_d$ superfield
doublet.  As a result, the neutralinos and neutrinos mix:
\bea
\lag_{\mN} &=& -\frac{1}{2}(\nu_i, -i\widetilde{B}, -i\widetilde{W}^3, \tilde{h}_d, \tilde{h}_u) \mN \left(\begin{array}{c}\nu_j \\ -i\widetilde{\mathcal{B}}\\ -i\widetilde{\mathcal{W}}^3\\ \tilde{h}_d\\ \tilde{h}_u \end{array} \right)\;. \nonumber \\
\eea
In the above expression, $\mN$ is a $7\times 7$ mass matrix.  As we are interested in models
with a strong hierarchy between the mass scales of the neutralinos and
the neutrinos, it is convenient to write $\mN$ as
\bea 
\mN &=& \left( \begin{array}{cc} m_{\nu} & m
  \\ m^T & \mathcal{M}_{\chi^0}
\end{array} \right),
\eea 
where $m_{\nu}$ is the $3\times 3$ mass matrix in the neutrino
sector and $\mathcal{M}_ {\chi^0}$ is the $4\times 4$ mass
matrix in the neutralino sector.  $m$ denotes the $3\times 4$ mixing matrix
which arises through R--parity violation.  An effective neutrino mass
matrix $\mnueff$ can then be defined via the see--saw mechanism
\bea \label{eq:mnu_eff}
\mnueff &\equiv& m_{\nu} - m \mathcal{M}_{\chi^0}^{-1} m^T.
\eea
At tree--level, in which $m_{\nu}=0$, it is given by
\cite{Joshipura:1994ib,Nowakowski:1995dx}
\newline
\begin{eqnarray}
(\mnueff)^{\textrm{tree}}_{ij} &=& \frac{\mu (M_1g^2_2 + M_2g^2)}{2 v_u v_d(M_1g^2_2 + M_2g^2)- 2 \mu M_1M_2}\Delta_i\Delta_j\nonumber\\
&&
\label{eq:treemass}
\label{eq:treelevel}
\end{eqnarray}
where
\begin{eqnarray}
\Delta_i &\equiv&  v_i - v_d \frac{\kappa_i}{\mu}, \qquad i=1,2,3 \, .
\label{eq:Lambda}
\end{eqnarray}
Here $v_i$ and $v_d$ are vacuum expectation values (VEVs) of the
sneutrino and ($H_d$) higgs fields.  An effective neutrino mixing matrix $U_{\nu}$ can then be defined via
the relation
\bea
U_{\nu}^T\mnueff U_{\nu} &=& {\rm diag}[m_{\nu i}], \qquad i=1,2,3.
\eea
The rank--1 structure of $(\mnueff)^{\rm tree}$ leads to only one
non--zero neutrino mass.  In order to fit neutrino oscillation data,
which implies at least two massive neutrinos, higher order corrections
must be included.  In fact, these corrections must be sizable as the
mass ratio of the two heaviest neutrinos is of order one,
\textit{cf. } Sect.~\ref{subsec:neutrinodata}.

In this paper, we therefore include the full one--loop contributions
to the neutrino--neutralino sector.  Our calculation follows closely
that of Refs.~\cite{Dedes:2006ni,Allanach:2007qc}.  However we go
beyond their approximations by including also the 1--loop LNV
corrections to the VEVs $v_i$, $v_d$ and $v_u$.  This is discussed in
more detail in section \ref{subsec:numerical_preliminary}.

Beyond tree--level, the matrix $m_\nu$ is filled by the loop
contributions to the neutrino masses.  A good measure of this
loop scale is set by contributions from so--called
$\Lambda\Lambda$ loops (see Sec.~\ref{sec:benchmarkpoints}):
\barr 
\phantom{.}\hspace{-0.8cm}(m_{\nu}^{\mathbf{\Lambda\Lambda}} )_{ij} &
=& \sum_{k,n} \left(\lam_{ikn}\lam_{jnk} A^l_{kn} + n_c\lam'_{ikn}
\lam'_{jnk} A^d_{kn}\right),
\label{eq:lamlamloop}
\earr
where $n_c=3$ is the color factor, and
\barr 
\phantom{.}\hspace{-1cm}A^f_{kn} &=& \frac{1}{32 \pi^2} m_{f_k} \sin
2\tilde{\phi}_n^f \,\ln\left( \frac{m^2_{\tilde{f}_{1
     n}}}{m^2_{\tilde{f}_{2 n}}} \right)  \\
&\sim& \frac{1}{16 \pi^2} m_{f_k}m_{f_n}\frac{(A_0 - \mu \tanb)} { m^2_{\tilde{f}_{L n}} -
    m^2_{\tilde{f}_{R n}}}\ln\left( \frac{m^2_{\tilde{f}_{L n
     }}}{m^2_{\tilde{f}_{R n}}} \right)
\label{eq:loopfunction}
\earr 
are loop functions for the $k$th generation fermion and $n$th
generation sfermions.  Here $f=\ell,d$ denotes a charged lepton
or a down--like quark. $A^f_{kn}$ depends on the fermion mass
$m_{f_k}$, the mixing angle $\tilde{\phi}^f_n$ for the rotation of the
left-- and right-- handed sfermion current eigenstates to the two mass
eigenstates, and the sfermion masses $m_{\tilde{f}_{1n}}$ and
$m_{\tilde{f}_{2n}}$.  The approximate expression is valid when the
sfermion left--right mixing is small.

The charged lepton--chargino mass matrix $\mCharge$ can be treated in a
similar fashion.  In particular an effective charged lepton mass
matrix $\mlepeff$ as well as its corresponding charged lepton mixing
matrices $U_{\ell L(R)}$ can be defined, which rotate the
left--\,(right--)\,handed charged leptons.  Consistent with our
notation, $\mCharge$ is defined in the same way as in
Ref.~\cite{Allanach:2009bv}.  To an excellent approximation, the
charged lepton masses can be obtained by
\bea
U^{\dagger}_{\ell L}\mlepeff U_{\ell R} &=& {\rm
diag}[m_{\ell i}], \qquad i=1,2,3.
\eea
Finally, the observable PMNS mixing matrix $\pmns$ is defined to be
\bea
\pmns &=& U^T_{\ell L}U_{\nu}.
\eea
To obtain a complete one--loop description of the PMNS matrix,
one--loop corrections to $U_{\ell L}$ are needed. These
corrections are however tiny compared with current experimental
uncertainties on neutrino oscillation observables, hence we neglect
them in the rest of this paper.

\subsection{Choice of Flavor Basis}\label{sec:FlavourBasis}

Since experimentally only the PMNS and
Cabbibo--Kobayashi--Maskawa (CKM) ($\ckm=U^{\dagger}_{uL}U_
{dL}$) \cite{Cabibbo:1963yz,Kobayashi:1973fv} mixing matrices
are known, simplifying assumptions parameterizing (left and right
handed) quark and lepton flavor mixing matrices are needed.  Following
\softsusy~\cite{Allanach:2001kg,Allanach:2009bv}, our computation 
assumes left--right symmetric mixings in the quark sector, and we work
in a basis where the charged lepton Yukawa matrix is diagonal.

Since we neglect the tiny one--loop corrections to $U_{\ell L}$,
$\pmns$ is determined by the form of the effective neutrino
mixing $U_{\nu}$.  For the quark mixings, there are two extreme cases
that could be considered:
\begin{itemize}
\item up--type mixing: $\ckm=U^{\dagger}_{uL}$, and $U_{dL}=
\mathbf{1}$ 
\item down--type mixing: $\ckm=U_{dL}$.  and $U_{uL}=
\mathbf{1}$
\end{itemize}
In the first case, $Y_U$ is non-- and $Y_D$ is diagonal, whereas the
second case is reversed. The choice of mixing can have
significant impact on the required magnitude of the $\lam'_
{ijk}$ couplings at the unification scale, especially for the
case $j\neq k$.  This is because in our model the bilinear LNV
couplings, $\kap_i$, that enter the tree--level mass $(\mnueff)
^{\rm tree}$ via Eq.~(\ref{eq:Lambda}) are generated via
renormalization group evolution. For example, there are
contributions of the form
\barr
\frac{d\kap_i}{dt} &\propto& \mu\, \lam'_{ijk}\times (Y_D)_{jk}, 
\label{eq:RGEkappa}
\earr 
where $t=\ln(Q/\mu_0)$, with $Q$ the renormalization scale and $\mu_0$
an arbitrary reference scale.  We see that the relative index
structure of the non--vanishing R--parity violating and conserving
Yukawa couplings is essential for the resulting magnitude of
$\kap_i$.

For concreteness, in this paper we work in the flavor basis with
up--type mixing, unless stated otherwise.  In this basis, the
$\lambda'_{ijk}$ couplings which are off--diagonal in $j,k$ do not
contribute significantly to $\mnueff$ at tree--level, but could be used
as parameters to adjust loop level contributions when fitting the
data. Note that because $Y_E$ is always diagonal in our model,
$\lambda_{ijk}$ couplings for $i,j\neq k$ can be utilized in a similar
fashion.  The changes that appear for down--type mixing is discussed
in sect.~\ref{sec:ChangesInBP}.

\subsection{Choice of cSSM benchmark point}\label{sec:benchmarkpoints}

\begin{figure}[t!]
\begin{center}
\epsfig{figure=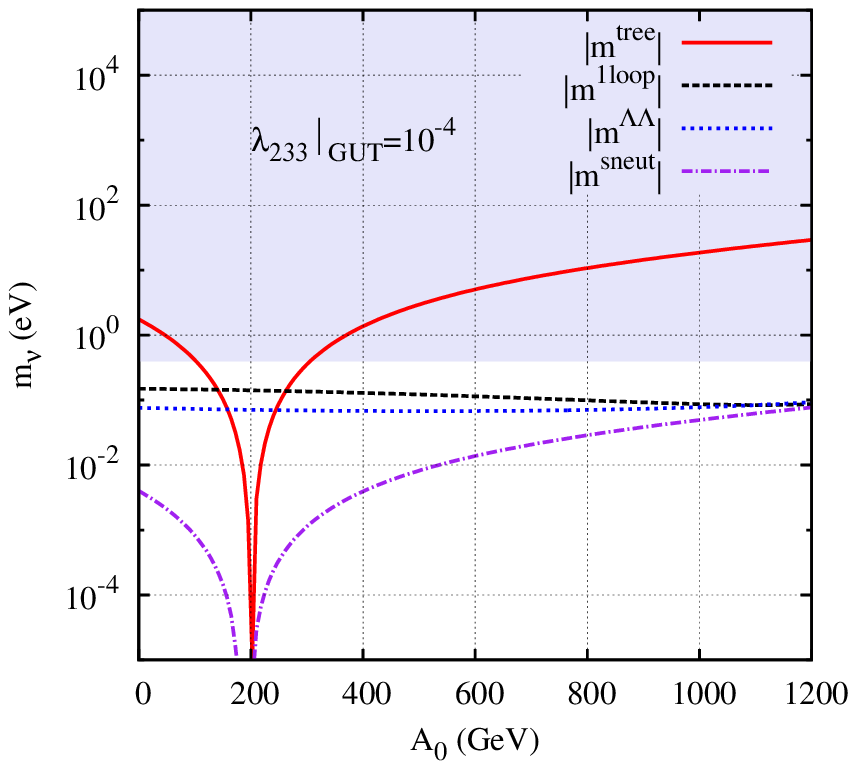,width=7.4cm}
\epsfig{figure=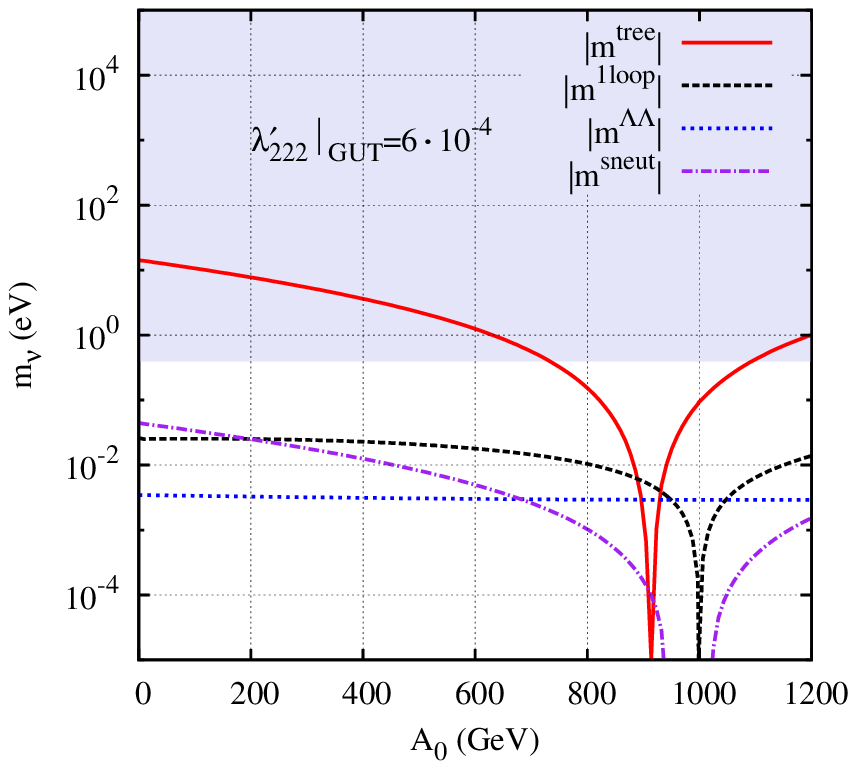,width=7.4cm}
\caption{$A_0$ dependence of the different contributions to the
  neutrino mass at the electroweak symmetry breaking scale for our
  benchmark point BP, with (top) $\lam_{233}|_{\mathrm {GUT}}= 10^{-4}$, (bottom)
  $\lamp_{222}|_{\rm GUT}= 6 \cdot 10^{-4}$. Note that only the
  absolute values of the contributions to the neutrino mass are
  displayed. The equations for $m_{\nu}^{\textrm{tree}}$ and
  $m_{\nu}^{\mathbf{\Lambda\Lambda}}$ are given in
  Eqs.~(\ref{eq:treemass}) and~(\ref{eq:lamlamloop}),
  respectively. $m_{\nu}^{\textrm{1loop}}$ represents the full
  one--loop corrections to the neutrino mass,
  $m_{\nu}^{\textrm{sneut}}$ represents the neutral scalar loops. The
  grey--shaded area is excluded by the cosmological bound. 
\label{fig:A0plot}}
\end{center}
\end{figure}

As has been noted in Ref.~\cite{Dreiner:2010ye}, there are preferred
regions of $B_3$ cSSM parameter space in which the neutrino
oscillation data can be more easily accommodated. This is illustrated
in Fig.~\ref{fig:A0plot} for one single LNV coupling. Recall that
there is only one tree--level neutrino mass, the second (and third)
neutrino mass scale is set by the 1--loop contributions
\cite{ref:footnote1}.  From Fig.~\ref{fig:A0plot} (a) [(b)] we see
that for a given $\lam$ [$\lam'$], in the parameter
region $100\lesssim \azero/{\rm GeV} \lesssim 300$ [$870
  \lesssim\azero/{\rm GeV} \lesssim 930$], the tree--level
  neutrino mass is sufficiently suppressed relative to the 1--loop
  neutrino mass to match the mild neutrino mass hierarchy required by the data of maximally 5.7, {\it cf.}
Eqs.~(\ref{eq:NH}), (\ref{eq:IH}). This region of parameter space
is determined by the fact that the tree--level neutrino mass
(solid cyan line in Fig.~\ref{fig:A0plot}) has a zero in $A_0$
parameter space due to RGE effects. This region exists for every $B_3$
cSSM parameter point, provided that
\bea A^{(\lamp)}_0 &\approx& 2\, M_{1/2} \label{eq:aolampminimum}\\ 
A^{(\lam)}_0 &\approx& \frac{M_{1/2}}{2}
\label{eq:aolamminimum}
\eea
for non--zero LNV couplings $\lam'_{ijk}$ or $\lam_{ijk}$,
respectively. Note that the position of the minimum is approximately
the same for all indices $i,j,k = 1,2,3$. Henceforth we denote the
$A_0$ minimum with respect to $\lam$ and $\lamp$ by $A^{(\lam)}_0$ and
$A^{(\lamp)}_0$ respectively. In this paper we focus on this
region; more details are given in
Sec.~\ref{subsec:minimizationprocedure}.  Therefore we have only 4
$\rp$--conserving parameters left, namely $\mhalf$, $\mo$, $\tanb$ and
$\sgnmu$.

\begin{table}[t]
\begin{center}
\begin{tabular}{|c|cccc|}
\hline
Particles & \multicolumn{4}{|c|}{Masses (GeV)}\\
\hline
$\tilde{g}$& 1146 & & &\\
$\tilde{\chi}^{\pm}_{1},\tilde{\chi}^{\pm}_{2}$&380 & 570&&\\
$\tilde{\chi}^{0}_{1},\tilde{\chi}^{0}_{2},\tilde{\chi}^{0}_{3},\tilde{\chi}^{0}_{4}$&\,\,204\,\,&\,\,380\,\,&\,\,552\,\,&\,\,571\,\,\\
$\tilde{u}_{1},\tilde{c}_{1},\tilde{t}_{1}$&1050&1050&1005&\\ 
$\tilde{u}_{2},\tilde{c}_{2},\tilde{t}_{2}$&1012&1012&858&\\ 
$\tilde{d}_{1},\tilde{s}_{1},\tilde{b}_{1}$&1053&1053&971&\\ 
$\tilde{d}_{2},\tilde{s}_{2},\tilde{b}_{2}$&1008&1008&1002&\\ 
$\tilde{e}_{1},\tilde{\mu}_{1},\tilde{\tau}_{1}$&353&353&346&\\ 
$\tilde{e}_{2},\tilde{\mu}_{2},\tilde{\tau}_{2}$&217&217&163&\\
$\tilde{\nu}_{e},\tilde{\nu}_{\mu},\tilde{\nu}_{\tau}$&343&343&331&\\
$h^0,A^0,H^0,H^{\pm}$&112&607&608&612\\
\hline
\end{tabular}
\caption{Mass spectrum of the benchmark point BP in the $\rp$
  conserving limit.  From top to bottom, the particles are the gluino,
  charginos, neutralinos, up--like squarks (2 rows), down--like
  squarks (2 rows), charged sleptons (2 rows), sneutrinos and the
  Higgses.  The charginos and neutralinos are ordered according to
  their masses.  For a scalar sparticle, a subscript 1(2) denotes that
  it is primarily `left'(`right') handed, \textit{i.e.} the
  superpartner of a left(right) chiral fermion.  This is the
  convention used in \texttt{SOFTSUSY}.  From left to right, the 4
  Higgses are the light CP--even Higgs, CP--odd Higgs, heavy CP--even
  Higgs and the charged Higgs.
\label{tab:BPspectrum}}
\end{center}
\end{table}

For easy comparison with Ref.~\cite{Dreiner:2010ye}, the benchmark point (BP) we use in this paper is chosen to be the same as Ref.~\cite{Dreiner:2010ye}:
\bea
\mhalf &=& 500 \textrm{ GeV} \nonumber \\
\mo &=& 100 \textrm{ GeV} \nonumber \\
\tanb &=& 20 \nonumber \\
\sgnmu &=& +1\,.
\eea
We have checked that this BP is
tachyon--free~\cite{Allanach:2003eb} and that the LEP2 exclusion bound
on the light SSM Higgs mass is fulfilled
\cite{Schael:2006cr,Barate:2003sz}.
The spectrum in the $\rp$ conserving limit is displayed in
Table~\ref{tab:BPspectrum}.  We see that the squark masses are of
order $\ord(1\,{\rm TeV})$, whereas the slepton masses are around 200--300
GeV.  The lightest supersymmetric particle (LSP) is a stau.  However
the presence of LNV couplings will render the LSP unstable, making
cosmological constraints on the nature of the LSP not applicable
\cite{Desch:2010gi,Dreiner:2008rv,Akeroyd:2001pm}.

It should also be pointed out that it is not possible to suppress
tree--level contributions for both $\lam$ and $\lam'$ simultaneously
for a universal $\azero$ parameter~\cite{Dreiner:2010ye}, as the two
minima do not coincide in the $A_0$ parameter space, \textit{cf.}
Eqs.~(\ref{eq:aolampminimum}), (\ref{eq:aolamminimum}). Therefore
scenarios such as those discussed in Ref.~\cite{Dey:2008ht}, where
there is no tree--level neutrino mass at all, are only possible in the
$B_3$ cSSM if there is only one type of LNV coupling, either $\lam$ or
$\lamp$.

It is also interesting to note that in the case of $\lam$ couplings
[Fig.~\ref{fig:A0plot} (a)], the full 1--loop contributions are well
approximated by the $\Lambda\Lambda$ loops, whereas in the case of
$\lamp$ couplings [Fig.~\ref{fig:A0plot} (b)], the approximation is
less satisfactory, and further 1--loop contributions such as
neutral scalar--neutralino loops also play an important role in parts
of the parameter space. However, around the $A^{(\lamp)}_0$ minimum, the
$\Lambda\Lambda$ loops still give a good order of magnitude estimate.

Note that viable neutrino masses could also be obtained away from the
$A_0$ minimum region by using only off--diagonal LNV couplings, since
the tree--level contribution is dominantly generated through diagonal
LNV couplings. Thus, scenarios involving only off--diagonal couplings
(and up--mixing if using $\lamp$ couplings) also lead to a suppression
of the tree--level contribution and could thus potentially reduce the
dependence on the $A_0$ minimum.

\subsection{Low--energy bounds on LNV parameters}\label{sec:lowEnergyBounds}

Once a set of $\Lambda$ couplings is specified to reproduce the
neutrino oscillation data, a natural question arises as to whether the
model is compatible with the large number of low energy observables
(LEOs).  If a considered model predicts LEO values close to current
experimental limits, future (non--)observations could (dis--)favor this
model.

An extended set of relevant bounds on LEOs is presented in
Refs.~\cite{Barger:1989rk,Allanach:1999ic,Barbier:2004ez}.  Typically
these constraints are more important for LNV couplings involving
lighter generations.  The reasons are two fold: Firstly, the fermion
mass term in the loop function $A^f_{kn}$ in
Eq.~(\ref{eq:loopfunction}) implies that, in order to generate a
neutrino mass contribution of the same size, LNV couplings involving a
light family index $k$ need to be much larger than corresponding
couplings with heavy family indices to compensate for the mass
suppression.  Secondly, experimental constraints generally provide
more stringent limits on LNV couplings involving light generations.

In the models presented in later sections, we compare our best fit
parameter values with the limits presented in
Ref.~\cite{Barbier:2004ez}, as well as a $\nubb$ bound on
$\lam'_{111}$ from Ref.~\cite{Allanach:2009xx}.  The bounds which are
most relevant for the discussion of our results are displayed below:
 
  \bit
\item[[b1]] $\mu\rightarrow eee$ decay:
\begin{eqnarray}
\lam_{nij} \lam_{n11} & \lesssim & 6.6\cdot 10^{-7}
\left( \frac{m_{\tilde{\nu}_n}}{100\,{\textrm GeV}} \right)^2,
\;\; i,j=12,21 \nonumber\\[2mm]
\lamp_{211}\lamp_{111} & \lesssim & 1.3 \cdot 10^{-4} \; \text{\cite{ref:footnote2}} \nonumber
\end{eqnarray}

\item[[b2]] $\mu-e$ conversion in nuclei: 
\begin{eqnarray}
\lam_{nij}\lam'_{n11} &\lesssim& 2.1\cdot
  10^{-8}\left(\frac{m_{\tilde{\nu}_n}}{100\,{\rm
      GeV}}\right)^2,\;\;\;  i,j=12,21 
\nonumber \\ 
\lam'_{2n1}\lam'_{1n1} &\lesssim& 4.3\cdot 10^{-8}\left(\frac{m_{\tilde{q}_n}}
{100\,{\rm GeV}}\right)^2,\;\;\;    n=2,3 
\nonumber\\ 
\lam'_{21n}\lam'_{11n}&\lesssim &4.5\cdot 10^{-8}\left(\frac{m_{\tilde{q}_n}}
{100\,{\rm GeV}}\right)^2,\;\;\; n=2,3  
\nonumber\\ 
\lam'_{211}\lam'_{111} & \lesssim& 4.3\cdot  10^{-8}\cdot \Delta^{-1} ,
\nonumber\\[3mm]
\Delta&\equiv& \left(\frac{100\,{\rm
      GeV}}{m_{\tilde{u}}}\right)^2-\left(\frac{2Z+N}{2N+Z}\frac{100\,{\rm
      GeV}}{m_{\tilde{d}}}\right)^2  \nonumber 
\end{eqnarray}
For $^{48}_{22}$Ti, $(2Z+N)/(2N+Z)=70/74$. This comes from the 
ratio of the number of valence up--quarks to that of the down--quarks
in a nuclei.  See Ref.~\cite{Kim:1997rr}.

\item[[b3]] $\mu$ decay:
\begin{equation}
\lam_{12k}\lesssim 0.08\left(\frac{m_{\tilde{e_{kR}}}}{100\,{\rm GeV}}\right)
\nonumber
\end{equation}
\item[[b4]] Leptonic $\tau$ decay:
\begin{equation}
\lam_{23k},\lam_{13k}\lesssim 0.08\left(\frac{m_{\tilde{e_{kR}}}}{100\,{\rm GeV}}\right)
\nonumber
\end{equation}

\item[[b5]] Forward--backward asymmetry of $Z$ decay:
\begin{eqnarray}
\lam_{i3k} (i\ne k \ne 3) &\lesssim& 0.25\left(\frac{m_{\tilde{\nu}_{\tau}}}
{100\,{\rm GeV}}\right) \nonumber \\
\lam_{i2k} (i\ne k \ne 2) &\lesssim& 0.11\left(\frac{m_{\tilde{\nu}_{\mu}}}
{100\,{\rm GeV}}\right) \nonumber 
\end{eqnarray}

\item[[b6]] Leptonic $K$--meson decay (here $i,j=12,21$):
\begin{eqnarray}
\lam_{n11}\lam'_{nij}
   &\lesssim& 1.0\cdot
  10^{-8}\left(\frac{m_{\tilde{\nu}_n}}{100\,{\rm
      GeV}}\right)^2,\nonumber\\ 
\lam_{n22}\lam'_{nij}  &\lesssim& 2.2\cdot
  10^{-7}\left(\frac{m_{\tilde{\nu}_n}}{100\,{\rm
      GeV}}\right)^2,\nonumber\\ 
\lam_{n12}\lam'_{nij}  &\lesssim& 6\cdot
  10^{-9}\left(\frac{m_{\tilde{\nu}_n}}{100\,{\rm
      GeV}}\right)^2,\nonumber\\ 
\lam_{n21}\lam'_{nij}  &\lesssim& 6\cdot
  10^{-9}\left(\frac{m_{\tilde{\nu}_n}}{100\,{\rm GeV}}\right)^2,\nonumber
\end{eqnarray}

\item[[b7]] $\mu \to e \gamma$:
\begin{eqnarray}
\lam_{nl2}\lam_{nl1}&<& 8.2\cdot 10^{-5}\cdot
\nonumber \\
&& \cdot\left[2\left(\frac{\rm 100\, GeV}{m_{\tilde{\nu}_L}}\right)^{2}
-\left(\frac{\rm 100\, GeV}{m_{\tilde{l}_L}}\right)^{2}\right]^{-1} 
\nonumber \\
\lam_{23n}\lam_{13n}&<&2.3\cdot 10^{-4} \cdot
\nonumber \\
&&\cdot\left[2\left(\frac{\rm 100\, 
GeV}{m_{\tilde{\nu}_L}}\right)^{2}-\left(\frac{\rm 100\, GeV}{m_
{\tilde{l}_R}}\right)^{2}\right]^{-1} 
\nonumber \\  
\lam'_{2nl}\lam'_{1nl} &<& 7.6\cdot 10^{-5}\left(\frac{m_{\tilde{d}_{lR}}}{\rm 100\, GeV}\right)^2,\;\; n=1,2\nonumber 
\end{eqnarray}

\item[[b8]] $\nubb$ (here $\tilde{f}=\tilde{e}_L,\tilde{u}_L,\tilde{d}_R$):
\begin{eqnarray}
|\lam'_{111}| \lesssim 5 \cdot 10^{-4} \Big(\frac{m_{\tilde
    f}}{100\textrm{GeV}}\Big)^{2} 
\Big(\frac{m_{{\tilde g}/{\tilde \chi}}}{100\textrm{GeV}}\Big)^{1/2}.  \nonumber
\end{eqnarray}
\eit 

These bounds are given in the mass basis, with the reference sparticle
mass scale set at 100 GeV. In order to compare our model values with
these bounds, we rotate to the mass basis and include the correct mass
dependence for all constraints derived from tree--level (4--fermion)
operators.

\section{Choice of LNV parameters}\label{sec:LNVpars}
In this section, we choose specific representative scenarios for the
LNV sector which will be used for the numerical fit of the neutrino
masses and mixings in Sec.~\ref{sec:numerical_analysis}.  First, as a
motivation to and a guide line in finding models, we discuss the
general neutrino mass matrix in the TBM approximation. As we have seen
in Sect.~\ref{subsec:neutrinodata}, this is a very good approximation to
the data.  Later, when performing our numerical fits, we use the
experimental values listed in Eqs.\,(\ref{eq:data1})--(\ref{eq:data3}).
In Sect.~\ref{diagScen} we limit the discussion to ``diagonal LNV
parameters" $\lam_{ijj}$ and $\lamp_{ijj}$. In
Section~\ref{offDiagScen} we discuss the more general case which
includes ``non--diagonal couplings", \textit{ i.e.}  $\lam_{ijk}$ and
$\lamp_{ijk}$ with $j \neq k$.

Since any LNV coupling $\lam_{ijk},\, \lamp_{ijk}$ could potentially
contribute to the effective neutrino mass matrix, we expect a large
number of possible solutions to
Eqs.~(\ref{eq:data1})--(\ref{eq:data5}).  It is well beyond the scope
of this paper to attempt to determine them completely. Instead we wish
to classify the types of solutions with a potentially minimal set of
parameters.  We thus make a series of simplifying assumptions,
restricting ourselves to a subset of couplings.  We will suggest 5
different scenarios (denoted \textbf{S1} to \textbf{S5}), each making
use of LNV coupling combinations from different types ($\lam$ and
$\lam'$) and generations, which we will make explicit as we proceed.

In order to obtain the neutrino mass matrix, we solve the 
equation
\beq
U_{TBM}^{\dagger} M^{TBM}_{\nu} U_{TBM} = {\rm diag}[m_{\nu \al}]\,,
\label{diagonalization}
\eeq
for $M^{TBM}_{\nu}$. Here the neutrino masses $m_{\nu \al} (\al =
1,2,3)$ fit the mass--squared differences and $U_{TBM}$ is given in
Eq.~(\ref{UTBM}).

It is natural to split up the resulting neutrino mass matrix into
three separate contributions, each of which is proportional to one
neutrino mass:\\

\begin{widetext}

\begin{eqnarray}
M^{TBM}_{\nu} &\equiv&  M_1 + M_2 + M_3 \nonumber \\
&=&\frac{m_{\nu 1}}{3} \left( \begin{array}{ccc}
\phantom{-}2 & -1 & \phantom{-}1\\
-1 & \phantom{-}1/2 & -1/2\\
\phantom{-}1 & -1/2 & \phantom{-}1/2 \end{array} \right)
+ \frac{m_{\nu 2}}{3} \left( \begin{array}{ccc}
\phantom{-}1 & \phantom{-}1 & -1\\
\phantom{-}1 & \phantom{-}1 & -1\\
-1 & -1 & \phantom{-}1  \end{array} \right)
+\frac{m_{\nu 3}}{2} \left( \begin{array}{ccc}
0 & 0 & 0\\
0 & 1 & 1\\
0 & 1 & 1  \end{array} \right) \\ [3mm]
&=& \frac{1}{6} \left( \begin{array}{ccc}
4m_{\nu 1} + 2m_{\nu 2} & \phantom{4m_{\nu 1}\,\,}2\alpha_{21} 
\phantom{+\,\,3\alpha_{31}} & \phantom{4m}-2\alpha_{21} \phantom
{+\,\,3\alpha_{31}} \\
\phantom{4m_{\nu 1}\,\,} \phantom{-}2\alpha_{21} & m_{\nu 1} + 2m_{\nu 2} 
+ 3m_{\nu 3} & \phantom{4m}-2\alpha_{21}+3\alpha_{31} \\
\phantom{4m_{\nu 1}}-2\alpha_{21} & \phantom{4m}-2\alpha_{21}+3
\alpha_{31} &  m_{\nu 1} + 2m_{\nu 2} + 3m_{\nu 3} \end{array} 
\right),
\label{eq:generalsplit} 
\end{eqnarray}
\end{widetext}
where the off--diagonal entries are written in terms of
\begin{eqnarray}
\alpha_{ij} &\equiv& \frac{\Delta m_{ij}^2}{m_{\nu i} + m_{\nu j}}\,.
\label{eq:generalsplit_1}
\end{eqnarray}

We observe that all three contributions $M_{\al}$ are of the
symmetric form
\beq
(M_{\al})_{ij} \propto c^{(\al)}_i c^{(\al)}_j \,. \label{eq:structure}
\eeq
If $U_{TBM}$ is orthogonal, this always follows from
Eq.~(\ref{diagonalization}), independent of its exact form. The
supersymmetric tree--level neutrino mass matrix displays an identical
structure if one assigns
\beq
c^{(tree)}_i \sim \lamp_{ijk} (Y_{D})_{jk}\,, \label{eq:ciLamp}
\eeq
or
\beq
c^{(tree)}_i \sim \lam_{ijk} (Y_{E})_{jk} \,. \label{eq:ciLam}
\eeq
This follows from a first--order approximation of
Eq.~(\ref{eq:treelevel}), making use of RGE considerations such as
Eq.~(\ref{eq:RGEkappa}) \cite{ref:footnote3}.  The dominant one--loop
level contribution to the neutrino mass matrix does not strictly
display the same structure, as can be seen from
Eq.~(\ref{eq:lamlamloop}).  However, for diagonal couplings ($j=k$),
one can make a similar assignment as in the tree--level case, 
\beq
c^{(loop)}_j \sim \lamp_{jkk} (m_{d})_{k} \label{eq:ciLoopLamp}
\eeq
or
\beq
c^{(loop)}_j \sim \lam_{jkk} (m_{\ell})_{k}  \,, \label{eq:ciLoopLam}
\eeq
\textit{cf. }  Eq.~(\ref{eq:loopfunction}). We discuss the
generalisation to non--diagonal couplings in Sect.~\ref{offDiagScen}.

For simplicity, we mainly focus on solutions which directly
reflect the form of Eq.~(\ref{eq:generalsplit}) (\textbf{S1} to
\textbf{S4})
\cite{ref:footnote6}, namely
\bea
c_1^{(1)}= -2 c_2^{(1)}= 2c_3^{(1)}=
\sqrt{\frac{2 m_{\nu_1}}{3}},\nonumber\\
c_1^{(2)} = c_2^{(2)} = -c_3^{(2)} = \sqrt{\frac{m_{\nu 2}}{3}},\nonumber\\
c_1^{(3)} = 0,\;\;\;  c_2^{(3)}=c_3^{(3)} = \sqrt{\frac{m_{\nu 3}}{2}} \;.\label{eq:simple_sol}
\eea

 This can minimally be achieved by allowing for
exactly one LNV parameter for each coefficient $c_i^{(\al)}$
\cite{ref:footnote5}.  The three matrices in
Eq.~(\ref{eq:generalsplit}) can then be described by 8 coefficients
\begin{equation}
\{c^{(1)}_{1,2,3},\;c^{(2)}_{1,2,3},\;c^{(3)}_{2,3}\}\,,
\end{equation}
where we have made use of the fact that $c^{(3)}_1=0$ in both the TBM
case {\it and the best--fit case}, under the assumption that
$\theta_{13}=0$. Since we need only two mass scales to describe the
neutrino data, we shall assume that the lightest neutrino is massless
in the NH and IH cases.  Depending on the scenario (NH, IH, DEG), we
thus need either five, six or eight non--zero coefficients
$c^{(\al)}_i$.

To illustrate possible alternatives, we show how ``non--diagonal''
couplings might contribute to neutrino masses in another example
(\textbf{S5}).

While we have presented the TBM approximation to display the general
coupling structure we are aiming for, in the numerical analysis below
we solve Eq.~(\ref{diagonalization}) not in the TBM approximation but
instead for the best--fit neutrino data given in
Eqs.~(\ref{eq:data1})--(\ref{eq:data5}). This results in slightly
different values for $c^{(i)}_j$. However, the deviation from the TBM
case is less than $7\%$ for each $c^{(i)}_j$.

\subsection{Diagonal LNV scenarios}\label{diagScen}
Scenarios involving only diagonal LNV couplings $\Lambda_{ijk}$ with
$j = k$ are the most straightforward to consider. With these we can
generate all neutrino mass matrix entries with a minimal set of LNV
couplings. The non--diagonal case requires additional
couplings, as we discuss below, \textit{cf.}~Sect.~\ref{offDiagScen}.
We first discuss normal hierarchy and inverted hierarchy scenarios and
then the degenerate case.

\bit
\item \textbf{Normal Hierarchy:} 

Since the first part of the neutrino mass matrix, $M_1$, is zero for
NH, we need only five LNV couplings to generate $M_\nu \equiv M_2 +
M_3$. In order to keep these two contributions $M_2$, $M_3$
(corresponding to the two non--zero neutrino mass eigenvalues) as
independent as possible, we use $\lam$ couplings for one and $\lamp$
couplings for the other matrix. If we now choose $A_0$ such that it
lies in the minimum region for either $\lam$ or $\lamp$ (we denote
this by $A_0^{(\lam)}$ and $A_0^{(\lamp)}$ respectively),
\textit{cf.}~Sect.~\ref{sec:benchmarkpoints}, we can generate one
neutrino mass eigenvalue at tree--level and one at loop--level in a
nearly independent fashion. This implies that the mass scales can be
easily adjusted. We focus on the case $A_0^{(\lamp)} \sim 2 \,\mhalf$,
where the contribution from $\lam'$ couplings to the tree--level mass
matrix is suppressed, because as we will show, for the IH scenarios
only this choice of $A_0$ is possible.  We briefly mention changes for
the case $A_0^{(\lam)} \sim \mhalf /2$ in NH scenarios during the
discussion in Sect. IV D.

Motivated by the observation that the first row/column of $M_3$ is
zero (\textit{i.e.} $c^{(3)}_1 =0$), and also $\lam_{111}=0$ due to
antisymmetry, we fit
\beq
(M_3)_{ij} \sim \lam_{i11} \lam_{j11} \;,
\eeq
(\textit{i.e.} $c^{(3)}_i \sim \lam_{i11}$). We then
automatically obtain the structure of $M_3$. Because we have chosen
$A_0^{(\lamp)}\sim 2 \mhalf$, this matrix is dominated by the
tree--level contribution.  In order to generate $M_2$ independently of
$M_3$ (at one--loop level), we choose
\beq
(M_2)_{ij} \sim \lamp_{ikk} \lamp_{jkk} \;,\label{eq:NH2}
\eeq
where $k$ is fixed. We present all three cases $k=1,2,3$ in
Table~\ref{tab:diagScenarios}, denoted \textbf{S1}, \textbf{S2} and
\textbf{S3}, respectively.

Additionally, we present one further scenario where we depart from the
correspondence $c^{(\al)}_i \sim \Lambda_i$. The motivation
for this is to consider a neutrino scenario where third generation
couplings are dominant, in analogy to the hierarchy of the SM
Yukawa couplings.  This scenario is particularly interesting because
it represents a lower limit on the required size of the LNV
couplings under the assumption that no further mechanism exists to
contribute to the neutrino masses. We discuss this aspect in
more detail in section~\ref{subsec:discussion}.  In order to be able
to fit the matrices $M_2$, $M_3$ only with third generation couplings
$\lam_{i33}$ and $\lamp_{i33}$, one of those matrices needs to fulfill
$(M_{i})_{3k} = 0$ due to the antisymmetry of $\lam$ in the
first two indices. To achieve this, we build a suitable superposition
of the matrices $M_2$ and $M_3$. We denote the new coefficients by
$\tilde c^{(\al)}_i$ in \textbf{S4} of Table~\ref{tab:diagScenarios}.

\item \textbf{Inverse Hierarchy:}

As mentioned in the case of Normal Hierarchy, $\lam_{ijj}$ couplings
will always lead to one row/column of zeros in the generated neutrino
mass matrix. Since in the case of Inverse Hierarchy, the two non--zero
matrices $M_1$ and $M_2$ are both non--zero in all entries, we take
this as motivation to fit $M_1$ and $M_2$ with $\lamp$ couplings only
(however, for completeness we also present one scenario with both
$\lam$ and $\lamp$ couplings, \textit{cf.} next paragraph). With only
$\lamp$ couplings present, we set the value of $A_0$ to $A_0^{(\lamp)}
\sim 2 \,M_{1/2}$, such that all tree--level contributions are
suppressed, and the two mass scales are both generated at loop level.
Otherwise the neutrino mass hierarchy would be much larger than
experimentally observed, \textit{cf.}  Sec.~\ref{sec:benchmarkpoints}.
We display the three possibilities arising from \bea (M_1)_{ij} &\sim&
\lamp_{ikk}\lamp_{jkk}\,,\\ (M_2)_{ij} &\sim& \lamp_{ill}
\lamp_{jll}\,, \label{eq:IH1} \eea where $l < k$ \cite{ref:footnote7}
in Table~\ref{tab:diagScenarios}.  These models are labelled (IH)
\textbf{S1}, \textbf{S2} and \textbf{S3}.

If we choose $\lam_{i\ell\ell}$ couplings instead of $\lamp_{i\ell
\ell}$ in Eq.~(\ref{eq:IH1}), this would again generate a (unwanted) 
row/column of zeros in $M_2$. Therefore, in this case we need to
combine, for example, $\lam_{i33}$ with $\lam_{322}$ in order to
generate non--zero entries for the third row/column of $M_2$.  Such a
combination of couplings generates a matrix of the form $c^ {(2)}_i
c^{(2)}_j$, where $c^{(2)}_{1, 2}$ and $c^{(2)}_{3}$ originate from
$\lam_{i33}$ and $\lam_{322}$ at tree--level respectively, because
these couplings generate $\kappa_i$ via the RGEs, \textit{cf.} 
Eqs.~(\ref{eq:treelevel}) and~(\ref{eq:RGEkappa}). In order to ensure
that $M_2$ is generated at tree--level, we still set $A_0^ {(\lamp)}=2
M_{1/2}$, such that we are able to fit Eq.~(\ref{eq:generalsplit}).
This case is also listed under \textbf{ S4} in
Table~\ref{tab:diagScenarios}.

\item \textbf{Degenerate Masses:} 

Since for degenerate masses, all three matrices $M_ {1,2,3}$ are
non--zero and of similar magnitude, this scenario is a combination of
choices made for NH and IH. As explained for the case of NH, we choose
\beq 
(M_3)_{ij} \sim \lam_{i11} \lam_{j11} \;.
\eeq 
To generate $M_1$ and $M_2$, we fit in analogy to the IH
case
 \bea (M_1)_{ij} \sim \lamp_{ikk} \lamp_{jkk}\\ (M_2)_{ij} \sim
\lamp_{ill} \lamp_{jll} \;. \eea 
These models are listed in Table~\ref{tab:diagScenarios} as (DEG)
\textbf{S1}, \textbf{S2} and \textbf{S3}. Here, as in the IH case, 
only the parameter choice $A_0^{(\lamp)}$ is possible in order to
suppress the $\lamp$ contribution to the tree--level neutrino mass.

\eit

\begin{table*}[t]
\begin{center}
\begin{tabular}{|c|l|l|l|}
\hline
& Normal Hierarchy (NH) & Inverse Hierarchy (IH) & Degenerate (DEG)\\
\hline
\textbf{S1} & $c^{(1)}_i \sim 0$ & $c^{(1)}_i \sim \lamp_{i11} $  &  $c^{(1)}_i \sim \lamp_{i11} $  \\
 &  $c^{(2)}_i \sim \lamp_{i11}$ &  $c^{(2)}_i \sim \lamp_{i22} $ & $c^{(2)}_i \sim \lamp_{i22} $\\
& $ c^{(3)}_i \sim \lam_{i11}$ & $c^{(3)}_i \sim 0$ & $ c^{(3)}_i \sim \lam_{i11}$ \\
\hline
\textbf{S2} & $c^{(1)}_i \sim 0$ & $c^{(1)}_i \sim \lamp_{i11} $ & $c^{(1)}_i \sim \lamp_{i11} $\\
 &  $c^{(2)}_i \sim \lamp_{i22}$ & $c^{(2)}_i \sim \lamp_{i33} $  & $c^{(2)}_i \sim \lamp_{i33} $\\
& $ c^{(3)}_i \sim \lam_{i11}$ & $c^{(3)}_i \sim 0$ & $ c^{(3)}_i \sim \lam_{i11}$\\
\hline
\textbf{S3} & $c^{(1)}_i \sim 0$ & $c^{(1)}_i \sim \lamp_{i22} $ & $c^{(1)}_i \sim \lamp_{i22} $\\
 &  $c^{(2)}_i \sim \lamp_{i33}$ &  $c^{(2)}_i \sim \lamp_{i33} $ & $c^{(2)}_i \sim \lamp_{i33} $\\
& $ c^{(3)}_i \sim \lam_{i11}$ & $c^{(3)}_i \sim 0$ & $ c^{(3)}_i \sim \lam_{i11}$\\
\hline
\textbf{S4} & $c^{(1)}_i \sim 0$ & $c^{(1)}_i \sim \lamp_{i33} $ & \\
 &  $\tilde c^{(2)}_i \sim \lamp_{i33}$ &  $c^{(2)}_i \sim \lam_{i33} \; \& \; \lam_{322} $ & \qquad -- \qquad\\
& $\tilde c^{(3)}_i \sim \lam_{i33}$ & $c^{(3)}_i \sim 0$ & \\
\hline
\textbf{S5} & $c^{(1)}_i \sim 0$ &  &  $c^{(1)}_i \sim \lamp_{i33} $ \\
 &    $ c^{(2)}_i \sim \lamp_{i23}  \; \& \; \lamp_{i32}  $ &  \qquad -- \qquad & $c^{(2)}_i \sim \lam_{i33} \; \& \; \lam_{322} $ \\
&   $ c^{(3)}_i \sim \lam_{i11}$ & &  $c^{(3)}_i \sim  \lam_{231}  \; \& \; \lam_{213}  \; \& \; \lam_{312}    \; (\& \; \lam_{313})$\\
\hline
\end{tabular}
\caption{Overview of the ``diagonal" (\textbf{S1} -- \textbf{S4}) and 
``non--diagonal" (\textbf{S5}) scenarios used for our
numerical analysis.
\label{tab:diagScenarios}}
\end{center}
\end{table*}

\subsection{Non--diagonal LNV scenarios}\label{offDiagScen}

In this section, we depart from the diagonal coupling scenarios and
discuss the effects of introducing ``non--diagonal" couplings.

When allowing for non--diagonal LNV couplings $\lamp_{ikl}$ ($\lam_
{ikl}$), $l \neq k$, we generally need more couplings than in the
diagonal case. This is because at one--loop level
\cite{ref:footnote8}, neutrino masses are dominantly generated
proportional to $\lamp_{ikl} \lamp_{ilk}$ ($\lam_{ikl}
\lam_{ilk}$). Thus, the assignment of one LNV coupling to one
$c^{(\al)}_i$ parameter (Eq.~(\ref{eq:structure})) is not possible for
the part of the neutrino mass matrix generated at 1--loop
level. Instead, we require
\beq
c^{(\al)}_i c^{(\al)}_j \sim \frac{1}{2} \cdot (\lamp_{ikl}
\lamp_{jlk} + \lamp_{ilk} \lamp_{jkl}) \; (m_d)_k \; (m_d)_l 
\label{eq:offdiagcis}
\eeq
where $k$, $l$ are fix (similarly for $\lam$ couplings). This
effectively doubles the number of LNV parameters if we choose $k \neq
l$.  Phenomenologically, one can distinguish between two cases:
\bit
\item[(a)] $\lamp_{ikl} \approx \lamp_{ilk}$ (same order of magnitude)
\item[(b)] $ \lamp_{ikl} \gg \lamp_{ilk}$ or vice versa (strong hierarchy)
\eit
In the first case (a), the size of the couplings will not differ
significantly from the diagonal case. For illustrative purposes, we
will present numerical results for a non--diagonal scenario similar to
the \textbf{S3} NH example, which we list under \textbf{S5} NH in
Table~\ref{tab:diagScenarios}. Here, we take as starting values
$\lamp_{i23} = \lamp_{i32}$ and thus, a simplified form of
Eq.~(\ref{eq:offdiagcis}) is $c^{(2)}_i \sim \lamp_{i32}$, similar to
the assignment in the diagonal case.

In the latter case (b), the size of the couplings become very
different from those in the diagonal scenarios. In particular, some of
the couplings can become very large. This is potentially of great
interest experimentally. However, various low--energy bounds could
potentially be violated. This can be illustrated with the help of the
following example with degenerate neutrino masses, which we list under
\textbf{S5} DEG in Table~\ref{tab:diagScenarios}. Here, the first two
neutrino masses are generated as in the case of \textbf{S4} IH
(however, now for normal mass ordering): $M_2$ is generated at
tree--level via diagonal $\lam_{i33}$ and $\lam_{322}$ couplings, and
$M_1$ is generated at loop--level via $\lamp_{i33}$ couplings.
However, now we additionally generate $M_3$ at one--loop level via the
3 off--diagonal $\lam$ couplings $\lam_{231}$, $\lam_{213}$ and
$\lam_{312}$. The latter do not lead to tree--level neutrino masses
because the leptonic Higgs--Yukawa coupling is (nearly) diagonal and
thus the tree--level generating term $\lam_{ijk} (Y_E)_{jk}$ is
(practically) zero.  As we will see, the benchmark point we use leads
to a very large $\lam_{231}$ beyond the perturbativity limit. For this
reason, a different BP point, labelled as BP2, will be introduced for
this scenario in Sect.~\ref{sec:offdiagDiscussion}
\cite{ref:footnote15}.

To obtain a qualitative understanding of the relative size of
the couplings, first note that $\lam_{133}$ contributes to both $M_2$
and $M_3$ due to the antisymmetry, $\lam_{133}\equiv -\lam_{313}$.  We
choose the $A_0^{(\lam')}$ minimum, and thus generate $M_2$ at tree
level.  The value of $\lam_{133}$ is therefore fixed, and is forced to
be small due to its coupling with the large tau Yukawa coupling
$(Y_E)_{33}$.  The matrices $M_1$ and $M_3$ are then generated at loop
level.  The coupling product $\lam_{231}\lam_{313}=-\lam_{231}
\lam_{133}$ is responsible for generating $(M_3)_{23}$.  This implies 
that $\lam_{231}$ needs to be large in order to compensate for the
smallness of $\lam_{313}$.  When now fitting $(M_3)_{22}\sim\lam_{231}
\lam_{213}$, the large $\lam_{231}$ then leads to a hierarchically 
smaller $\lam_{213}$ in order to be consistent with the experimental
result.  Similarly, $\lam_{231}$ leads to a small $\lam_{312}$ by
their contribution to $(M_3)_{33}$ via $\lam_{312}\lam_{321}(A^{l}
_{12}+A^{l}_{21})$ as shown in Eq.~(\ref{eq:lamlamloop}).

\section{Numerical Analysis}
\label{sec:numerical_analysis}
In this section, we present the numerical results. We will first
discuss the relevant aspects of \softsusyv\ for our analysis. We then
describe our minimization procedure. Next we present our
best--fit solutions for the normal hierarchy, inverted hierarchy and
the degenerate case, respectively. In the last subsection we discuss
these results.

\subsection{Preliminaries}
\label{subsec:numerical_preliminary}
Our numerical simulation is performed using an adaptation of
\softsusyv\ , which will be made public in the near future. Until then, 
we refer interested readers to the \softsusy\, manual
\cite{Allanach:2009bv} for the detailed procedure of obtaining
the $\bt$ SSM mass spectrum.  We use the program package \minuit\, and
a Markov chain Monte Carlo method (Metropolis--Hastings algorithm) for
fitting the LNV couplings $\Lambda_{ijk}$ to the neutrino data.

We now comment briefly on the additional features we include in
\softsusy\, and the determination of the $\pmns$ mixing in the
following.  Our calculation improves on \softsusyv\ by including the
one--loop contributions to the neutrino--neutralino mass matrix, as well
as all $\rpv$ tadpole corrections to the Higgs and sneutrino VEVs.
Because the superfields $L_i$ and $H_d$ have the same quantum numbers,
we organize the computation to treat these fields on equal footing.
To ensure the accuracy, an independent calculation was performed
without using this symmetry.  We have also checked that in the $\rp$
conserving limit our results agree with the internal results in
\softsusyv. 

The $\rpv$ tadpole corrections are included in the
\softsusy\ iteration procedure which minimizes the 5--dimensional EW
symmetry breaking neutral scalar potential.  The effective 3 $\times$
3 neutrino mass matrix $\mnueff$ and the effective neutrino mixing
matrix $U_{\nu}$ are calculated at the EWSB scale given an input set
of LNV parameters at the unification scale.  Note that within
\softsusy\,, the condition that the charged lepton mixing matrix is
diagonal is imposed at the electroweak scale. Thus, $\pmns = U_{\nu}$
\cite{ref:footnote10}.

\subsection{Minimization Procedure}
\label{subsec:minimizationprocedure}
Our goal is to find numerical values for each LNV scenario specified
in Table~\ref{tab:diagScenarios}, such that we obtain the
experimentally observed neutrino data,
Eqs.~(\ref{eq:data1})--(\ref{eq:data5}), at the 1$\,\sigma$ level by
means of least--square fitting.  In order to achieve this also in
degenerate scenarios, which necessarily involve some fine--tuning (as
we discuss in Sect.~\ref{subsec:discussion}), we use a multistep
procedure as outlined below.

We take as initial values for each set of LNV parameters at the
unification scale $M_X$
\beq
\Lambda_{ikk}  \sim c^{(\al)}_i \frac{1}
{{(Y_f)_{kk}}} \label{eq:ansatz}
\eeq 
(no summation over $k$) as specified in
Table~\ref{tab:diagScenarios}. $f$ denotes a down quark for a
$\lam'$ and a charged lepton for a $\lam$ coupling.  The
proportionality factor is estimated from the upper bound on the LNV
couplings which comes from the upper bound on the neutrino mass from
WMAP measurements, \textit{cf.} Ref.~\cite{Dreiner:2010ye}.

Next, we perform a pre--iteration within our modified version of
\softsusy, where we make the simplifying assumption that the
generation of the tree--level (by $\Lambda = \lam$) and 1--loop level
(by $\Lambda=\lamp$) neutrino mass matrices $M_{\alpha}$ in
Eq.~~(\ref{eq:generalsplit}) are independent of each other. So
for each $M_\alpha$ we separately fit the relevant $\Lambda_{ijk}$. In
our iteration procedure we set
\beq
\Lambda_{ijk}|_{new} = \sqrt{\frac{(M_{\al}^{obs})_{ii}}{(M_{\al}^
{softsusy})_{ii}}} \Lambda_{ijk}|_{old}\,. \label{iteration}
\eeq
Here $M_{\al} ^{softsusy}$ is the effective neutrino mass matrix (at
1--loop level) obtained via the seesaw--mechanism with \softsusy\,. 
In the first step we use the initial values corresponding to 
Eq.~(\ref{eq:ansatz}). We obtain $(M_{\al}^{obs})_{ii}$ by 
inverting Eq.~(\ref{diagonalization}), without using the TBM
approximation. For $m_{\nu\alpha}$ we use the experimental best--fit
values. And for the diagonalization matrix $U$, we implement the
general form, using $\theta_{12},\,\theta_{23}$ from the experimental
best--fit, as well as $\theta_{13}=0$. In Eq.~(\ref{iteration}) 
there is also no sum over $i$.

This gives a very good order of magnitude estimate for all LNV
couplings and thus a suitable starting point for our least--square
fit. However, so far each set of couplings $\Lambda_{ijk} \sim
c^{(\al)}_i/(Y_f)_{kk}$ has only been fit separately for each $\al$,
while keeping the other LNV couplings equal to zero. When fitting all
LNV couplings simultaneously, they can affect each other via the RGEs
and through contributions to the other $M_{\al}^{obs}$.  Note that
these effects are easily controllable for NH and IH scenarios.
However, in the case of DEG scenarios, some
strong cancellations occur for some entries of the
effective neutrino mass matrix, \textit{e.g.} the $(M_{\nu})_{13} =
(M_1)_{13} +(M_2)_{13} $ entry in Eq.~(\ref{eq:generalsplit}).  Here,
both individual entries $(M_{\al})_{13} $ are of the order of the
generated neutrino mass, but the resulting $(M_{\nu})_{13} $ entry is
at least 3 orders of magnitude smaller. This will become relevant in
the next step of our procedure.

After these first approximations, we next fit all LNV
parameters specified for each scenario in
Table~\ref{tab:diagScenarios} simultaneously. We calculate the full
$7\times7$ neutralino--neutrino mass matrix with
\softsusy. The $3\times3$ neutrino mass matrix is then obtained via
the seesaw mechanism, and is used in order to extract
predictions for the neutrino masses and mixing angles.

We define a $\chi^2$ function 
\beq
\chi^2 \equiv \frac{1}{N_{obs}}\sum_{i=1}^{N_{obs}}  \left(
\frac{f^{softsusy}_i - f^{obs}_i}    {\delta_i}  \right)^2 
\label{eq:chi2}
\eeq
where $f_i^{obs}$ are the central values of the $N_{obs}$ experimental
observables defined in Eqs.~(\ref{eq:data1})--(\ref{eq:data5}),
$f_i^{softsusy}$ are the corresponding numerical predictions and
$\delta_i$ are the 1$\,\sigma$ experimental uncertainties. We
minimize Eq.~(\ref{eq:chi2}) with a stepping method of the program
package \minuit\ for the NH/IH case. In the DEG scenarios,
\minuit\ initially does not converge due to the points made in the
last paragraph. Therefore, we first use the Hastings--Metropolis
algorithm to obtain a $\chi^2 < \ord(10)$. Subsequently, the same
\minuit\ routine as in the NH/IH case is used. We accept a
minimization result as successful if our minimization procedure yields
$\chi^2 < 1$.

Simultaneously, we ensure that the conditions
\barr
\sum_i m_{\nu_i} \lesssim 0.4 \textrm{ eV} \nonumber \\
\sin^2(\theta_{13}) < 0.047
\earr
are fulfilled.

\section{Discussion of Results}\label{subsec:discussion}

\begin{table*}[t]
\begin{center}
\begin{tabular}{|c|l|l|l|}
\hline
& Normal Hierarchy & Inverse Hierarchy & Degenerate\\
\hline
\textbf{S1} 		& &{\vtext{$\lamp_{111} =  \:\:\:\,\mathbf{3.94 \cdot 10^{-2}}$ [b1],[b2],[b8]}} & {\vtext{$\lamp_{111} = \:\:\:\,\vmtext{5.85 \cdot 10^{-2}}$  [b1],[b2],[b7],[b8]}} \\
			&   &\vtext{$\lamp_{211} =\vmtext{-1.88 \cdot 10^{-2}}$ [b1],[b2]} & {\vtext{$\lamp_{211} = \vmtext{-3.63 \cdot 10^{-2}}$  [b1],[b2],[b6],[b7]}}\\
			&  & $\lamp_{311} = \:\:\:\,1.94 \cdot 10^{-2}$ & \vtext{$\lamp_{311} = \:\:\:\,\vmtext{3.35 \cdot 10^{-2}}$ [b6]}\\
&\vtext{$\lamp_{111} =  \:\:\:\,\vmtext{1.12 \cdot 10^{-2}}$ [b2],[b8]}   & $\lamp_{122} =\:\:\:\,1.21 \cdot 10^{-3} $ & $\lamp_{122} =  \:\:\:\,2.18 \cdot 10^{-3} $ \\
&\vtext{$\lamp_{211} =  \:\:\:\,\vmtext{8.76 \cdot 10^{-3}}$ } & $\lamp_{222} = \:\:\:\,1.27 \cdot 10^{-3}$ & $\lamp_{222} = \:\:\:\,1.63 \cdot 10^{-3}  $ \\
&  $\lamp_{311} =  -1.48 \cdot 10^{-2}$ & $\lamp_{322} =  -1.31 \cdot 10^{-3}$ & $\lamp_{322} = -2.09\cdot 10^{-3} $ \\
&  \vtext{$\lam_{211} = \:\:\:\,\vmtext{1.52\cdot 10^{-2}}$ [b2]}  & & {\vtext{$\lam_{211} =  \:\:\:\,\vmtext{2.55 \cdot 10^{-2}}$  [b2],[b6]}} \\
& $\lam_{311} = \:\:\:\,1.37\cdot 10^{-2}$ &  & \vtext{$\lam_{311} = \:\:\:\, \vmtext{2.28 \cdot 10^{-2} }$ [b6]}\\
\hline

\textbf{S2} 		&  & \vtext{$\lamp_{111} = \:\:\:\,\vmtext{3.99\cdot 10^{-2}}$  [b1],[b2],[b8]} &  \vtext{$\lamp_{111} =\:\:\:\,\vmtext{6.87\cdot 10^{-2}}$ [b1],[b2],[b7],[b8]}   \\
			&  & \vtext{$\lamp_{211} =\vmtext{-1.81\cdot 10^{-2}}$ [b1],[b2]} & \vtext{$\lamp_{211} = \vmtext{-2.90\cdot 10^{-2}}$ [b1],[b2],[b6],[b7]}\\
			&  & $\lamp_{311} =\:\:\:\,1.89\cdot 10^{-2}$ & \vtext{$\lamp_{311} = \:\:\:\,\vmtext{3.18\cdot 10^{-2}} $ [b6]} \\
			&  $\lamp_{122} =  \:\:\:\,5.08 \cdot 10^{-4}$ & $\lamp_{133} = \:\:\:\, 3.09\cdot 10^{-5}$ & $\lamp_{133} = \:\:\:\,4.99 \cdot 10^{-5}$ \\
			& $\lamp_{222} =  \:\:\:\,3.88 \cdot 10^{-4}$ & $\lamp_{233} = \:\:\:\, 3.21\cdot 10^{-5}$ & $\lamp_{233} = \:\:\:\,2.98 \cdot 10^{-5}$ \\
			&$\lamp_{322} =  -6.97 \cdot 10^{-4}$  & $\lamp_{333} = -3.35		\cdot 10^{-5}$ & $\lamp_{333} = -7.43 \cdot 10^{-5}$ \\
			& $\lam_{211} = \:\:\:\,1.52\cdot 10^{-2}$ & & \vtext{$\lam_{211} = \:\:\:\,\vmtext{2.99 \cdot 10^{-2}}$ [b2],[b6]}\\
			& $\lam_{311} = \:\:\:\,1.37\cdot 10^{-2}$ &  & \vtext{$\lam_{311} =  \:\:\:\,\vmtext{2.10\cdot 10^{-2}}$ [b6]}\\
\hline
\textbf{S3} 		&   & $\lamp_{122} =  \:\:\:\, 1.80\cdot 10^{-3}$ &  $\lamp_{122} =\:\:\:\,  2.93\cdot 10^{-3}$  \\
			& &  $\lamp_{222} =  -8.29\cdot 10^{-4}$ & $\lamp_{222} = -1.98\cdot 10^{-3}$\\
			&  & $\lamp_{322} =  \:\:\:\, 8.64\cdot 10^{-4}$ & $\lamp_{322} =  \:\:\:\, 5.79\cdot 10^{-4}$ \\
			& $\lamp_{133} =  \:\:\:\, 1.30\cdot 10^{-5}$ & $\lamp_{133} =  \:\:\:\, 3.11\cdot 10^{-5}$ & $\lamp_{133} = \:\:\:\, 5.18\cdot 10^{-5}$ \\
			&   $\lamp_{233} =  \:\:\:\, 4.84\cdot 10^{-6}$ & $\lamp_{233} =  \:\:\:\, 3.22\cdot 10^{-5}$ & $\lamp_{233} = \:\:\:\, 5.78\cdot 10^{-5}$ \\
			&  $\lamp_{333} = -2.28\cdot 10^{-5}$& $\lamp_{333} = -3.32\cdot 10^{-5}$ & $\lamp_{333} = -5.13\cdot 10^{-5}$ \\
			& $\lam_{211} =  \:\:\:\,1.55\cdot 10^{-2}$ & & $\lam_{211} = \:\:\:\, 1.71\cdot 10^{-2}$ \\
			& $\lam_{311} =  \:\:\:\,1.40\cdot 10^{-2}$ &  & $\lam_{311} = \:\:\:\, 3.08\cdot 10^{-2}$ \\
\hline

\textbf{S4} 		& $\lamp_{133} =  - 6.80\cdot 10^{-6}$  & $\lamp_{133} = \:\:\:\, 3.96\cdot 10^{-5}$ &    \\
			&  $\lamp_{233} =   \:\:\:\, 2.81\cdot 10^{-5}$ &  $\lamp_{233} =  -2.81\cdot 10^{-5}$ & \\
			& $\lamp_{333} =   \:\:\:\,4.21\cdot 10^{-5}$ & $\lamp_{333} = \:\:\:\, 2.89 \cdot 10^{-5}$ &  \\
			& $\lam_{133} =  \:\:\:\,1.32\cdot 10^{-6}$ &  $\lam_{133} =  \:\:\:\,3.23\cdot 10^{-6}$ &  \\
			& $\lam_{233} =  \:\:\:\,2.70\cdot 10^{-6}$ &   $\lam_{233} =  \:\:\:\, 3.48\cdot 10^{-6}$ & \\
									&  & $\lam_{322} =  - 5.64\cdot 10^{-5}$ &\\
\hline
\textbf{S5} 		& $\lamp_{123} =   \:\:\:\, 5.76\cdot 10^{-5}$  &  &$\lamp_{133} =    -3.11\cdot 10^{-5}$     \\
			&  $\lamp_{132} =   \:\:\:\, 5.75\cdot 10^{-5}$ & &  $\lamp_{233} =  \:\:\:\,8.79\cdot 10^{-5}$  \\
			& $\lamp_{223} =   \:\:\:\,6.23\cdot 10^{-5}$ & & $\lamp_{333} =  -4.14 \cdot 10^{-5}$   \\
			& $\lamp_{232} =   \:\:\:\,6.24\cdot 10^{-5}$  &  &$\lam_{133} =  \:\:\:\, 1.99\cdot 10^{-6}$     \\
			&  $\lamp_{323} =   -5.88\cdot 10^{-5}$ & &  $\lam_{233} =  \:\:\:\,4.08\cdot 10^{-6}$  \\
			& $\lamp_{332} =   -6.00\cdot 10^{-5}$ & & $\lam_{322} =   -2.57 \cdot 10^{-5}$   \\
			& $\lam_{211} =  \:\:\:\,1.52\cdot 10^{-2}$ &  & $\lam_{231} =  -5.67 \cdot 10^{-2}$  \\
			& $\lam_{311} =  \:\:\:\,1.39\cdot 10^{-2}$ & &   $\lam_{213} =  -2.03\cdot 10^{-5}$  \\
			&						&  & $\lam_{312} =  \:\:\:\,2.54\cdot 10^{-3}$ \\
\hline
\end{tabular}
\caption{Best--fit points for the LNV parameters at the unification
  scale $M_X$ for our benchmark point BP and $A_0^{(\lamp)} = 912.3$
  GeV, except for {\bf S5} DEG, where BP2 and $A_0^{(\lamp)} =
  1059.2$ GeV are used, \textit{cf.}
  Sect.~\ref{sec:offdiagDiscussion}. The couplings printed in bold
  violate one of the low--energy bounds [b1]--[b7] which are listed in
  Sect.~\ref{sec:lowEnergyBounds}. Note that the values are given at 2
  significance level only for better readability. In order to
  reproduce the results, higher significance is needed as is clear
  from Eq.~(\ref{eq:w}). Readers are encouraged to contact the authors
  to obtain the exact values. }
\label{tab:results}
\end{center}
\end{table*}

\begin{table*}[t]
\begin{center}
\begin{tabular}{|c|ccccrrr|ccccrrr|ccccrrr|}
\hline
&\multicolumn{7}{|l|}{Normal Hierarchy} 
&\multicolumn{7}{|l|}{Inverse Hierarchy} 
&\multicolumn{7}{|l|}{Degenerate}\\
\hline
Data 
&           &           &         & &   &     &
&$c^{(1)}_1:$&$c^{(1)}_2:$&$c^{(1)}_3$&=& 2.09 : & $-$0.98 :   &1 
&$c^{(1)}_1:$&$c^{(1)}_2:$&$c^{(1)}_3$&=& 2.09 : & $-$0.98 :   &1  \\
&$c^{(2)}_1:$&$c^{(2)}_2:$&$c^{(2)}_3$&=& 0.94 : & 0.99 :      &$-$1 
&$c^{(2)}_1:$&$c^{(2)}_2:$&$c^{(2)}_3$&=& 0.94 : & 0.99 :      &$-$1 
&$c^{(2)}_1:$&$c^{(2)}_2:$&$c^{(2)}_3$&=& 0.94 : & 0.99 :      &$-$1 \\
&           &$c^{(3)}_2:$&$c^{(3)}_3$&=&        & 0.99 :      &1 
& & & && & & 
&           &$c^{(3)}_2:$&$c^{(3)}_3$&=&        & 0.99 :      & 1 \\ 
\hline
\hline
\textbf{S1} 
&               &               &              & &       &       &
&$\lamp_{111}:$&$\lamp_{211}:$&$\lamp_{311}$&=& 2.04 : & $-$0.97 : &1 
&$\lamp_{111}:$&$\lamp_{211}:$&$\lamp_{311}$&=& 1.75 : & $-$1.09 : &1 \\
&$\lamp_{111}:$&$\lamp_{211}:$&$\lamp_{311}$&=& 0.75 : & 0.59 :    &$-$1 
&$\lamp_{122}:$&$\lamp_{222}:$&$\lamp_{322}$&=& 0.93 : & 0.97 :    &$-$1 
&$\lamp_{111}:$&$\lamp_{211}:$&$\lamp_{311}$&=& 1.04 : & 0.78 :    &$-$1 \\
&               &$\lam_{211}:$&$\lam_{311}$&=&       & 1.11 :    &1
&               &               &              & &       &          &
&               &$\lam_{211}:$&$\lam_{311}$&=&       & 1.19 :    &1 \\
\hline
\textbf{S2} 
&               &               &              & &        &          &
&$\lamp_{111}:$&$\lamp_{211}:$&$\lamp_{311}$&=& 2.12 : &$-$0.96 : & 1 
&$\lamp_{111}:$&$\lamp_{211}:$&$\lamp_{311}$&=& 2.11 : &$-$0.91 : & 1 \\
&$\lamp_{122}:$&$\lamp_{222}:$&$\lamp_{322}$&=& 0.73 : & 0.56 :   &$-$1 
&$\lamp_{133}:$&$\lamp_{233}:$&$\lamp_{333}$&=& 0.93 : & 0.96 :   &$-$1 
&$\lamp_{133}:$&$\lamp_{233}:$&$\lamp_{333}$&=& 0.67 : & 0.40 :   &$-$1 \\
&               &$\lam_{211}:$&$\lam_{311}$&=&        & 1.11 :   & 1
&               &            &           & &        &          &
&               &$\lam_{211}:$&$\lam_{311}$&=&        & 1.42 :   & 1 \\
\hline
\textbf{S3} 
&             &             &            & &        &           &
&$\lamp_{122}:$&$\lamp_{222}:$&$\lamp_{322}$&=& 2.09 : & $-$0.96 : &1 
&$\lamp_{122}:$&$\lamp_{222}:$&$\lamp_{322}$&=& 5.06 : & $-$3.41 : &1 \\
&$\lamp_{133}:$&$\lamp_{233}:$&$\lamp_{333}$&=& 0.57 : & 0.21 :    &$-$1
&$\lamp_{133}:$&$\lamp_{233}:$&$\lamp_{333}$&=& 0.93 : & 0.97 :    &$-$1 
&$\lamp_{133}:$&$\lamp_{233}:$&$\lamp_{333}$&=& 1.01 : & 1.13 :    &$-$1 \\
&             &$\lam_{211}:$&$\lam_{311}$&=&        & 1.11 :    &1
&             &            &            & &        &           &
&             &$\lam_{211}:$&$\lam_{311}$&=&        & 0.56 :    &1 \\
\hline
\end{tabular}
\caption{Ratios of the LNV parameters at the unification scale $M_X$
  for scenarios \textbf{S1, S2} and \textbf{S3} and the ratios
  $c^{(\alpha)}_1:c^{(\alpha)}_2:c^{(\alpha)}_3$ inferred from
  experimental data.  For comparison, the ratios
  $c^{(\alpha)}_1:c^{(\alpha)}_2:c^{(\alpha)}_3$ in the TBM limit are
  $(2:-1:1)$, $(1:1:-1)$ and $(0:1:1)$ for $\alpha=1,2$ and $3$
  respectively.}
\label{tab:results_ratio}
\end{center}
\end{table*}


We present our numerical results in Table~\ref{tab:results}. In the
three columns, we show our best--fit solutions for normal hierarchy,
inverse hierarchy and degenerate masses, respectively. In the five
rows, we show our solutions for the various scenarios enlisted in
Table~\ref{tab:diagScenarios}. \textbf{S1}--\textbf{S4} are the
``diagonal'' LNV scenarios, while \textbf{S5} involves non--diagonal
couplings, as discussed in the previous section.  In order to
illustrate the low energy bounds most relevant to our scenarios, we
also display models which do not satisfy all constraints.  These
solutions are highlighted in bold and the violated bound(s) are also
stated.

\subsection{Diagonal LNV Scenarios}

We first discuss some general features of the best fit parameter sets.
Focusing on the three scenarios \textbf{S1}--\textbf{S3}, some ratios
among the LNV couplings are displayed in
Table~\ref{tab:results_ratio}.  We see that the results reflect the
basic structure of our ans\"atz Eq.~(\ref{eq:ansatz}). In particular,
the relative signs among different LNV couplings are reproduced.
However, the relative magnitude among the couplings are expected to
deviate somewhat from Eqs.~(\ref{eq:ansatz}) and
~(\ref{eq:simple_sol}). One reason is that our LNV couplings should
mirror the structure of Eq.~(\ref{eq:simple_sol}) at the electroweak
scale, while in Table~\ref{tab:results} and
Table~\ref{tab:results_ratio} the couplings are given at the
unification scale. So RG running needs to be taken into
account. However the change in the LNV couplings when going to the
unification scale is not uniform for all couplings. Also, we fit the
oscillation data given in Sec.~\ref{sec:neutrinodata} instead of the
TBM approximation, such that the $c_i^{(\alpha)}$ differ from
Eq.~(\ref{eq:simple_sol}) already by up to $7\%$ percent.

We also see from Table~\ref{tab:results_ratio} that the LNV parameters
in the IH scenarios follow the pattern of $c_i^{(\alpha)}$ more
closely than those in the NH and DEG scenarios. For the IH scenarios,
the tree level contribution is suppressed by choosing $A_0$
appropriately.  The neutrino mass matrix entries are dominated by loop
contributions and the associated couplings should then reflect the
near TBM structure as well as the orthogonality of the vectors
$c^{(\alpha)}$.  However for the NH and DEG scenarios, the significant
contributions from both tree and loop masses mean that while the
$c_i^{(\alpha)}$ have the expected ratios for each $\alpha$ after
pre--iteration, once contributions from different $\alpha$'s are
combined for the full iteration they interfere with each other. For
example, the presence of $\lam$ couplings changes the position of the
$A_0^{(\lam')}$ minimum, making the contributions of the $\lam'$
couplings to the tree level masses less suppressed, thus leading to
the larger deviation.

It is clear from Eq.~(\ref{eq:ansatz}) that the magnitude of diagonal
LNV couplings should decrease from first to third generation (while
generating the same neutrino masses), because the LNV couplings have
to balance out the effect of the Higgs--Yukawa--couplings, which
increase with generation. For example, comparing the size of
$\lamp_{ikk}$ in scenarios \textbf{S1}--\textbf{S3} in the IH case,
one observes that the difference in magnitude of the LNV couplings
mirrors the hierarchy of down--type quark masses, $\lamp_{ijj} /
\lamp_{ikk} \sim (m_d)_k / (m_d)_j$ for fixed index $i$.

As we see in Table~\ref{tab:results}, models involving first
generation couplings ($\lamp_{111}$ and $\lamp_{211}$) are disfavored
due to strong constraints from $\mu\rightarrow eee$ [b1], $\mu$--$e$
conversions [b2] and $\nubb$ [b8].  In addition, the $\lam_{211}$ in
\textbf{S1} NH, \textbf{S1} DEG and \textbf{S2} DEG violate the
two--coupling bound from $\mu$--$e$ conversion [b2] in conjunction
with the large $\lamp_{111}$ coupling.  Limits on leptonic K--meson
decay [b6] and $\mu\to e\gamma$ [b7] are also seen to be violated in
degenerate scenarios \textbf{S1} DEG and \textbf{S2} DEG involving
diagonal first generation couplings.  The second generation LNV Yukawa
couplings are of the order of $10^{-3}$ ($10^{-4}$) for IH and DEG
(NH) scenarios \cite{ref:footnote9} and safely satisfy all low--energy
bounds.  The third generation couplings take on values between
$10^{-5}$ and $10^{-6}$.

Collider implications of the solutions we obtained will be discussed
in section \ref{sec:Collider}.  Generally speaking, the stringent low
energy bounds on the first generation couplings could be evaded in
models with heavier supersymmetric mass spectra. In these models the
relatively large couplings could still lead to interesting collider
phenomenology, for example resonant production of sparticles
\cite{Dreiner:2000vf}.  These couplings could also have significant
impact on the RG running of the sparticle masses, and result in
observable changes to the sparticle spectrum when compared with those
in the $\rp$--conserving limit.  In particular, new LSP candidates may
be obtained even within the $\bt$ cSSM framework
\cite{Dreiner:2008ca}.

In contrast, third generation couplings are tiny, \textit{e.g.} the
\textbf{S4} NH model in Table~\ref{tab:results}.  However these small
couplings could result in a finite decay length for the LSP and hence
potential detection of displaced vertices in a collider.  See
Ref.\,\cite{Allanach:2007qc} for numerical estimates.

\begin{figure}[t!]
\begin{center}
\epsfig{figure=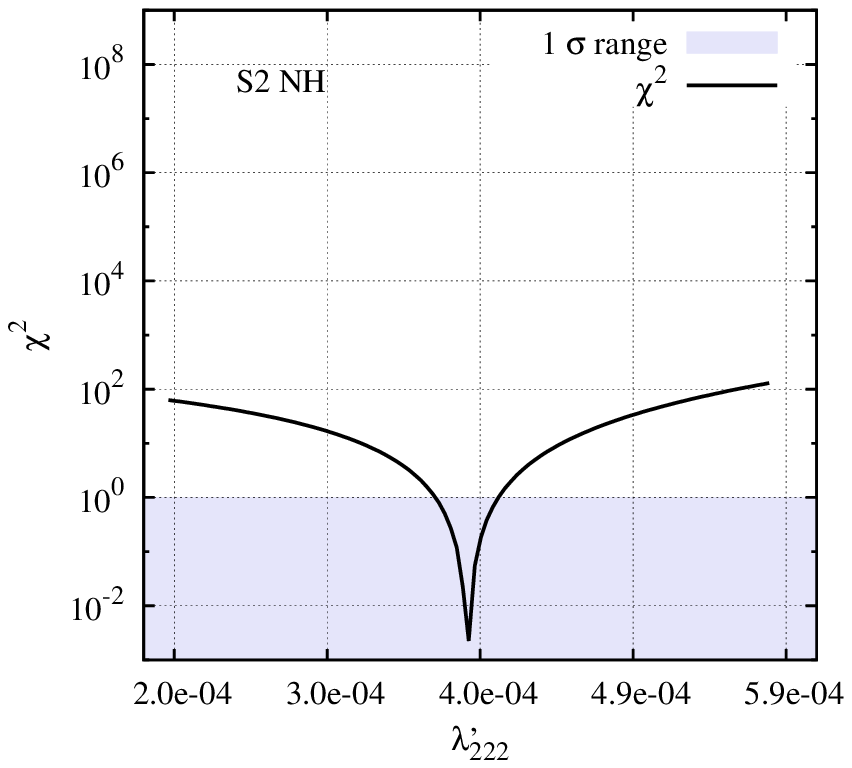,width=7.4cm}
\epsfig{figure=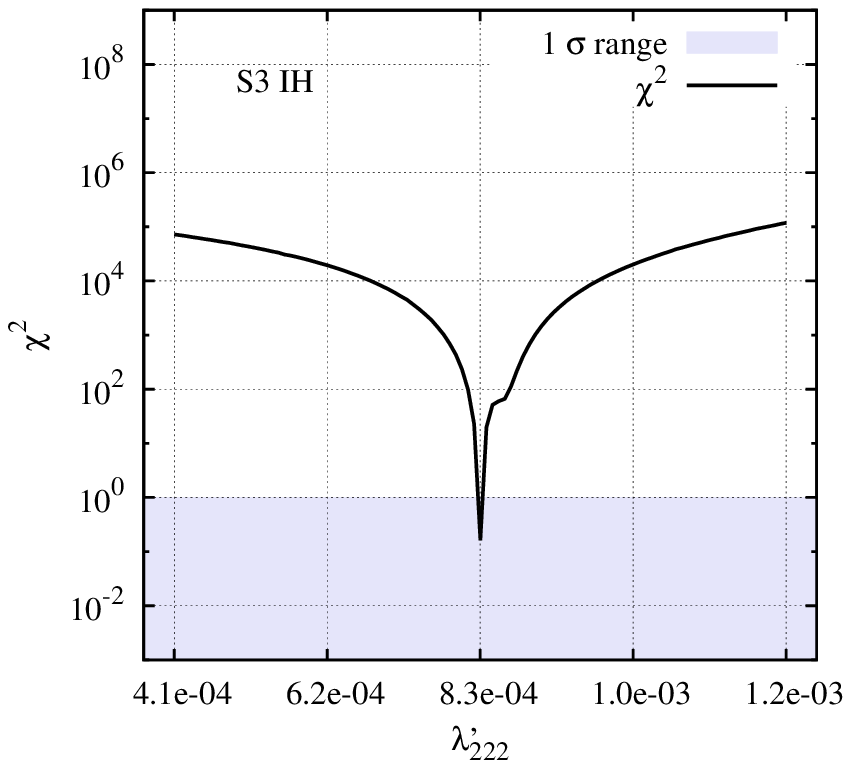,width=7.4cm}
\epsfig{figure=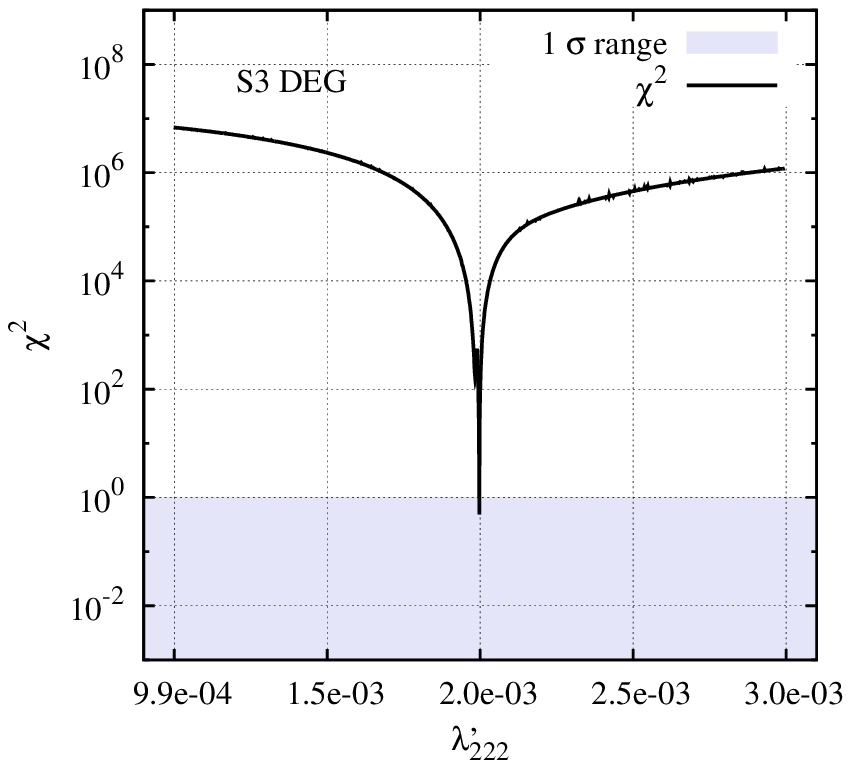,width=7.4cm}
\caption{Variation of $\chi^2$ as a function of $\lamp_{222}$ for
  scenarios {\bf S2} NH, {\bf S3} IH and {\bf S3} DEG. The glitches in
  {\bf S3} IH and {\bf S3} DEG are associated with the
  `crossing--over' of mass eigenstates when $\lam'_{222}$ is varied.
  See text for more discussion.
\label{fig:chi2plots}}
\end{center}
\end{figure}

In Fig.~\ref{fig:chi2plots}, we display the changes in $\chi^2$ for a
few selected scenarios ({\bf S2} NH, {\bf S3} IH and {\bf S3} DEG)
when a LNV coupling is varied within [0.5:1.5] times the best--fit
value.
We define a ``width'' for a $\chi^2$ minimum to be
\beq
w \,\equiv\, \frac{\Delta \Lambda|_{\chi^2<3}}{\Lambda|_
{\chi^2\approx 0}} \;,
\eeq
so that a large (small) $w$ value may be interpreted as less (more)
fine--tuning between different LNV couplings.

Clearly the NH case looks significantly better than the IH/DEG cases:
\bea
w(\mathrm{NH}, \Lambda =\lamp_{222}) &=& 1.1 \cdot 10^{-1}\;, \nonumber \\
w(\mathrm{IH},  \Lambda = \lamp_{222}) &=& 7.4  \cdot 10^{-3}\;,  \nonumber \\
w(\mathrm{DEG},  \Lambda = \lamp_{222}) &=& 4.8  \cdot 10^{-4}\;. \label{eq:w}
\eea
In fact, since the neutrino masses in our model are free parameters to
be fitted to the data, it is natural for these masses to be
non--degenerate.  To obtain the two (three) quasi--degenerate masses
in the IH (DEG) spectrum thus requires a certain amount of
fine--tuning, which should be reflected in the value of $w$.  Recall
from Eqs.~(\ref{eq:generalsplit}) and (\ref{eq:generalsplit_1}) that
due to a small (zero) $\sin[\theta_{13}]$ in the near (exact) TBM
limit, there are small off--diagonal entries for an inverted or
degenerate mass spectrum.  Specifically, $\alpha_{21}$ is small in
both cases, while $\alpha_{31}$ is also small in a DEG spectrum.  As a
result, there are small off--diagonal entries for both IH and DEG
scenarios but not for a normal hierarchy, while in our set--up the
diagonal and off--diagonal entries of $M_{\alpha}$ are of the same
order for each $\alpha$. Therefore, a way to understand this fine
tuning technically would be by considering the size of the
off--diagonal entries of $\mnueff$. We discuss the three cases
separately.

In the case of NH, the off--diagonal entries in $\mnueff$ will be of
the same order as the diagonal values. In this case, the experimental
observables are fairly insensitive to changes of up to $\ord(10\%)$ in
the LNV sector, \textit{cf.}  Eq.~(\ref{eq:w}).

For IH, we have two nearly degenerate mass eigenstates. Therefore, the
tree--level and the loop contribution have to be of the same order,
with a near--cancellation occurring between the off--diagonal entries
of $M_1$ and $M_2$. This results in a significantly larger width of
the $\chi^2$ minimum than in the NH case.

For the same reason, in the DEG cases even larger fine--tuning is
required in order to obtain \textit{three} nearly degenerate neutrino
masses.  Actually, in the limit $M \gg \Delta M \sim \Delta m^2/M$,
where $M$ is the mass scale of the heaviest neutrino, all
off--diagonal entries will have a magnitude of $\ord(\Delta M)$, and the
width $w$ can be approximated by
\begin{eqnarray}
  \Lambda^2 &\sim& M\,,  \\[3mm]
  \frac{\Delta \Lambda}{\Lambda} &\sim& \frac{1}{2}\frac{\Delta M}{M}.
\end{eqnarray}
A consequence of such fine--tuning is that if $\mnueff$ is deformed
slightly (for example due to changes in model parameters or technical
aspects such as low convergence threshold in the spectrum
calculation), the angles can change a lot since they are especially
sensitive to the (small) off--diagonal entries of $\mnueff$. In
contrast, the mass values are much more stable, with their sum
determined by the diagonal entries of $\mnueff$.

This can be illustrated by changing the implementation of the LNV
parameters in the numerical code from 6 significant figures to 3: the
masses change by less than 1 percent, whereas the angles change by a
factor of order one. Therefore the values displayed in
Table~\ref{tab:results}, especially those for the IH and DEG cases,
need to be taken with caution.  However, listing more digits would
result in worse readability, so we ask readers interested in
reproducing our results to contact the authors for more precise
values.

\begin{figure*}[ht!]
  \begin{center}
  \epsfig{figure=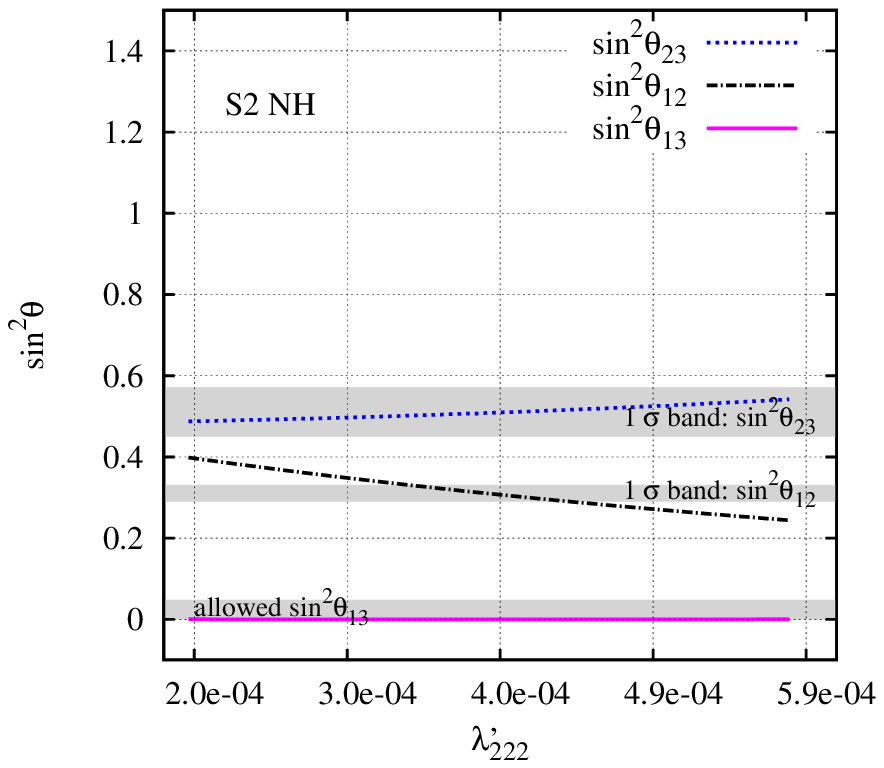,width=7.4cm}
  \epsfig{figure=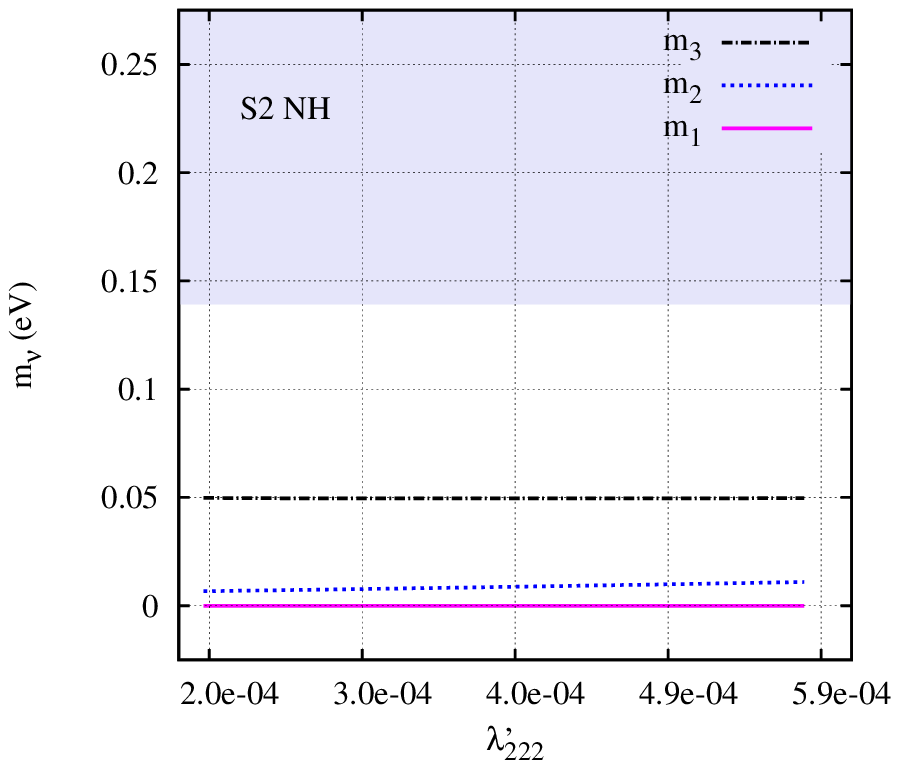,width=7.4cm}
  \epsfig{figure=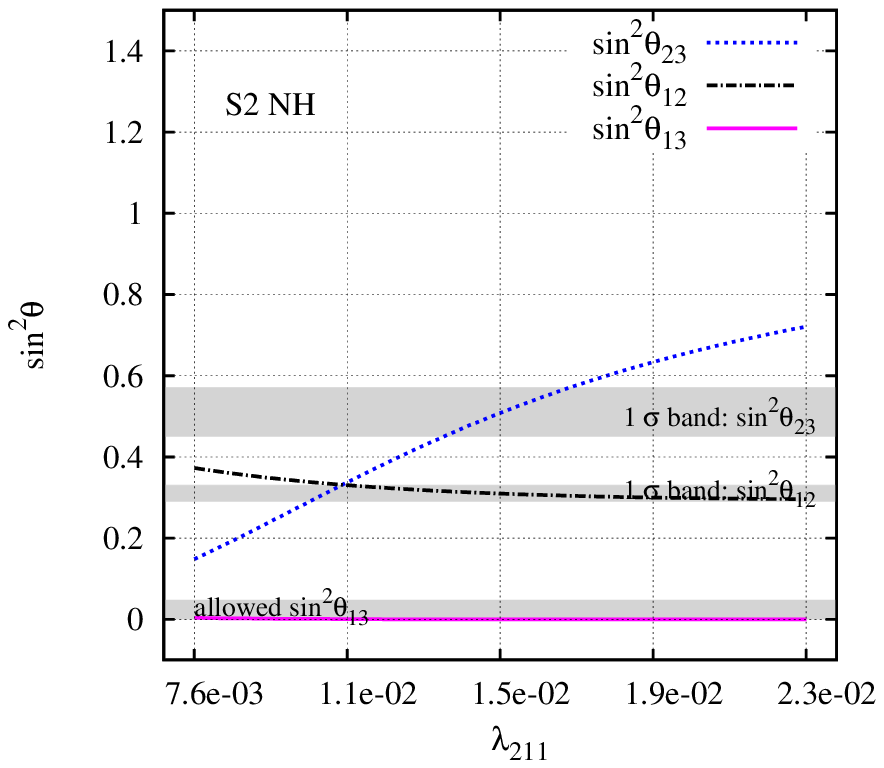,width=7.4cm}
  \epsfig{figure=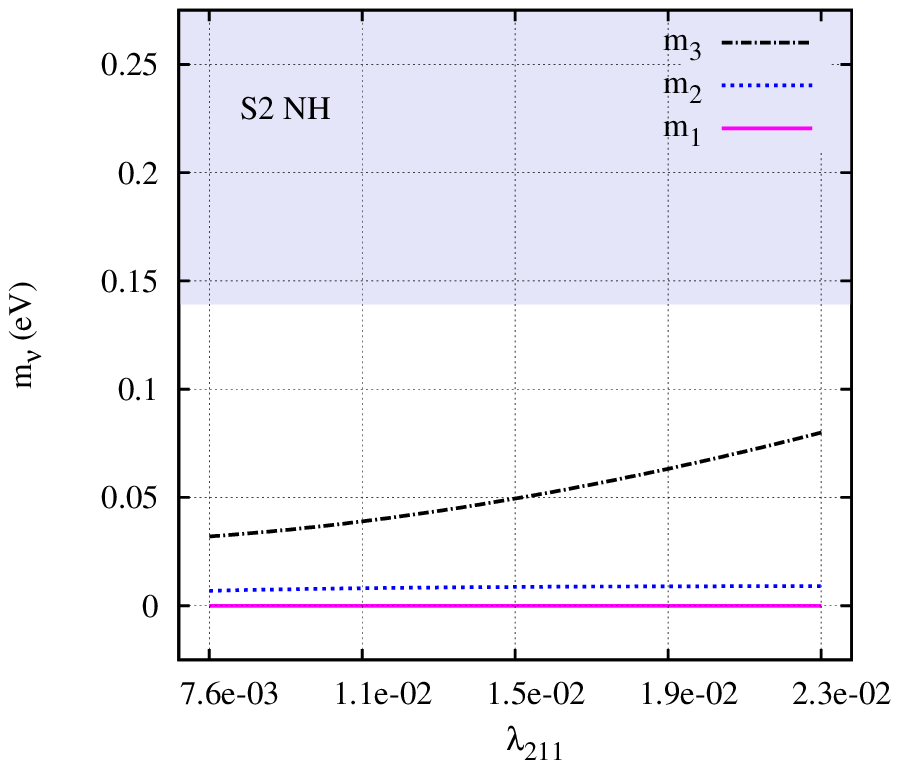,width=7.4cm}
  \caption{Variation of the mixing angles (left) and
  the mass eigenvalues (right) as functions of $\lamp_{222}$
  (top) and $\lam_{211}$ (bottom) for scenario {\bf S2} NH.  The best
  fit values for $\lamp_{222}$ and $\lam_{211}$ are located at
  the centre of the plots.  On the plots of mixing angles, the grey
  bands are experimentally viable regions for (from top to bottom)
  ${\rm sin}^2\theta_{23}$, ${\rm sin}^2\theta_{12}$ and ${\rm
  sin}^2\theta_{13}$.  On the plots of mass eigenvalues, values inside
  the grey bands are disfavoured by cosmological considerations.
  \label{fig:SCEN2NH}} \end{center}
\end{figure*}

\begin{figure*}[ht!]
  \begin{center}
  \epsfig{figure=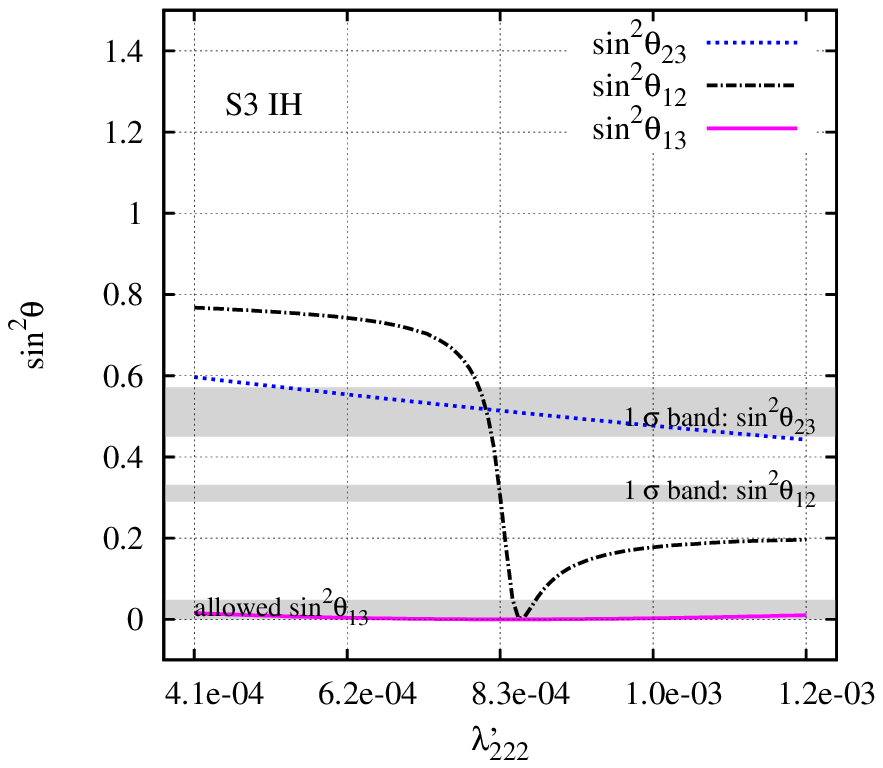,width=7.4cm}
  \epsfig{figure=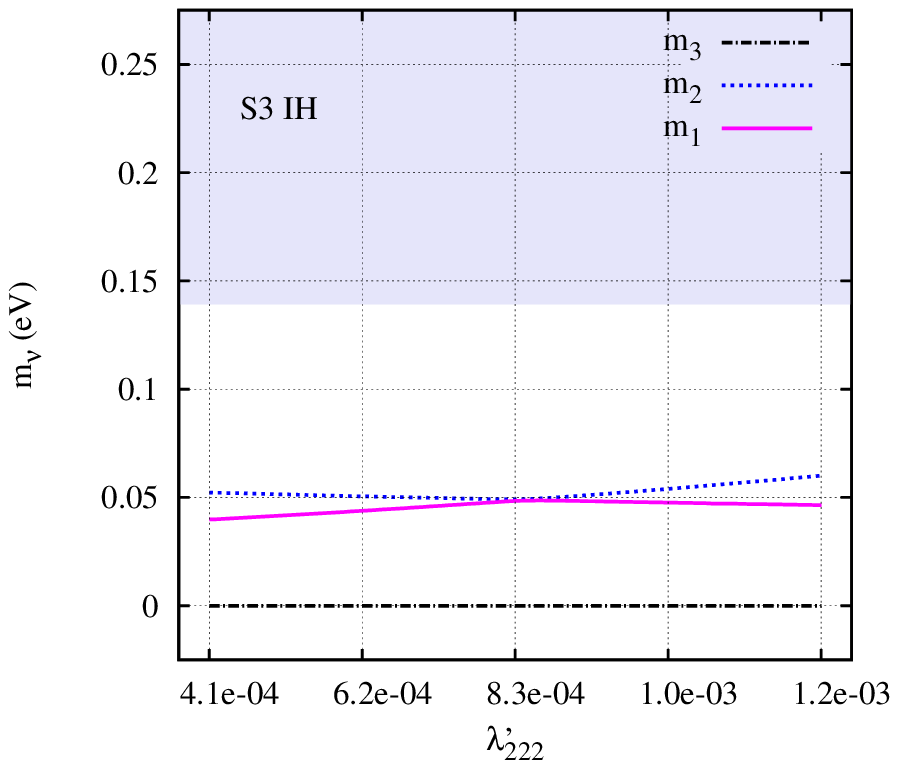,width=7.4cm}
  \epsfig{figure=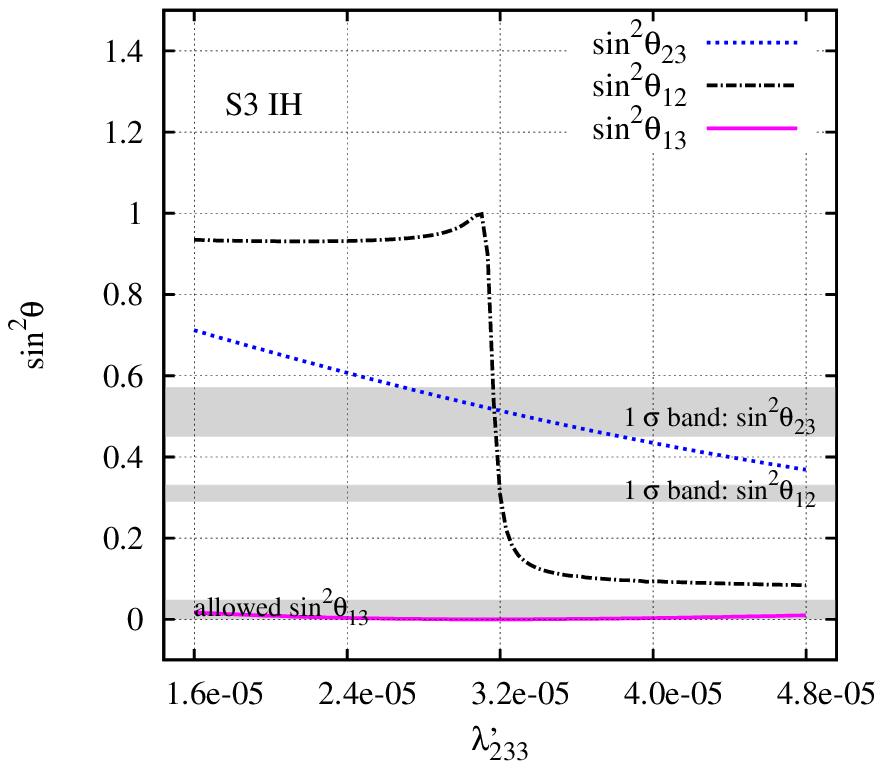,width=7.4cm}
  \epsfig{figure=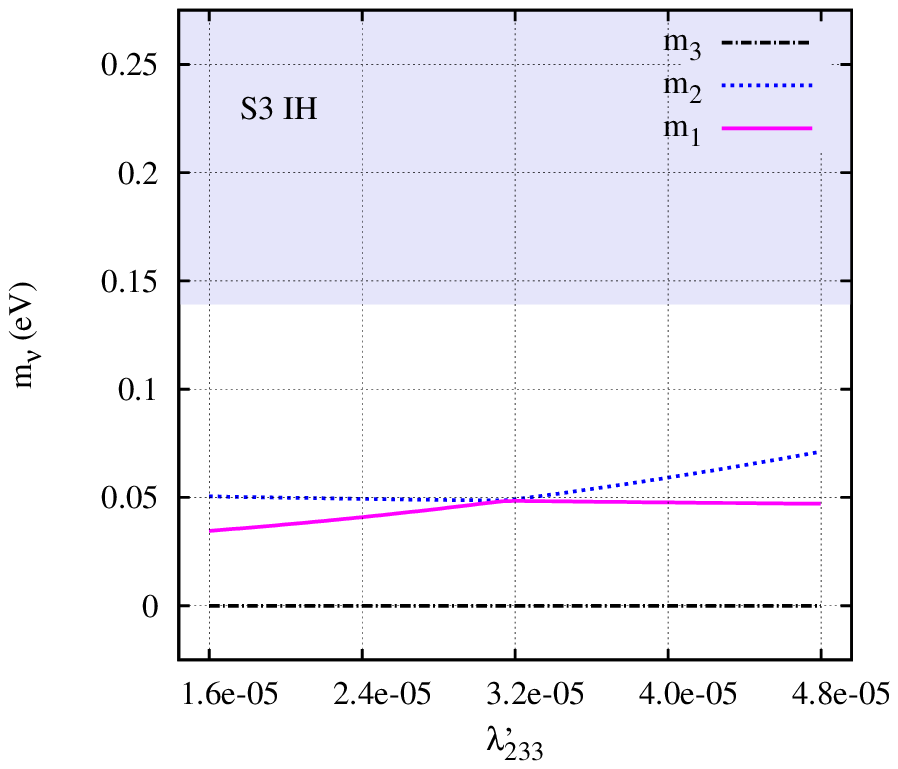,width=7.4cm}
  \caption{Variation of the mixing angles (left) and the mass
  eigenvalues (right) as functions of $\lamp_{222}$ (top) and
  $\lamp_{233}$ (bottom) for scenario {\bf S3} IH.  The best--fit
  values of $\lamp_{222}$ and $\lamp_{233}$ are located at the
  centre of the plots.  On the plots of mixing angles, the grey bands
  are experimentally viable regions for (from top to bottom) ${\rm
  sin}^2\theta_{23}$, ${\rm sin}^2\theta_{12}$ and ${\rm
  sin}^2\theta_{13}$.  On the plots of mass eigenvalues, values inside
  the grey bands are disfavoured by cosmological considerations.
  \label{fig:SCEN3IH}} 
  \end{center}
\end{figure*}

\begin{figure*}[ht!]
  \begin{center}
  \epsfig{figure=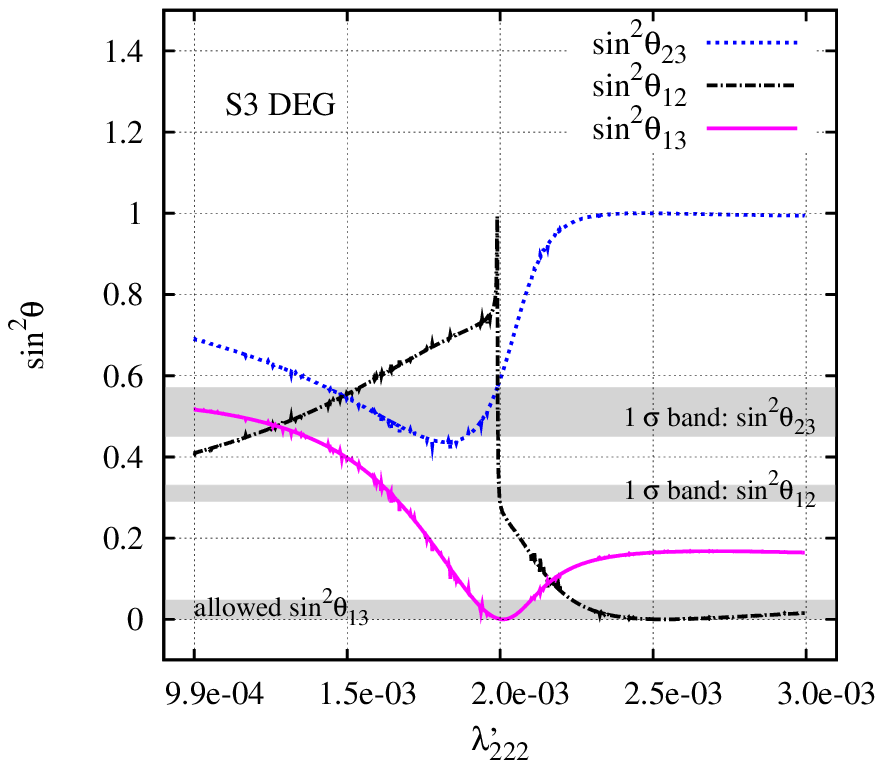,width=7.4cm}
  \epsfig{figure=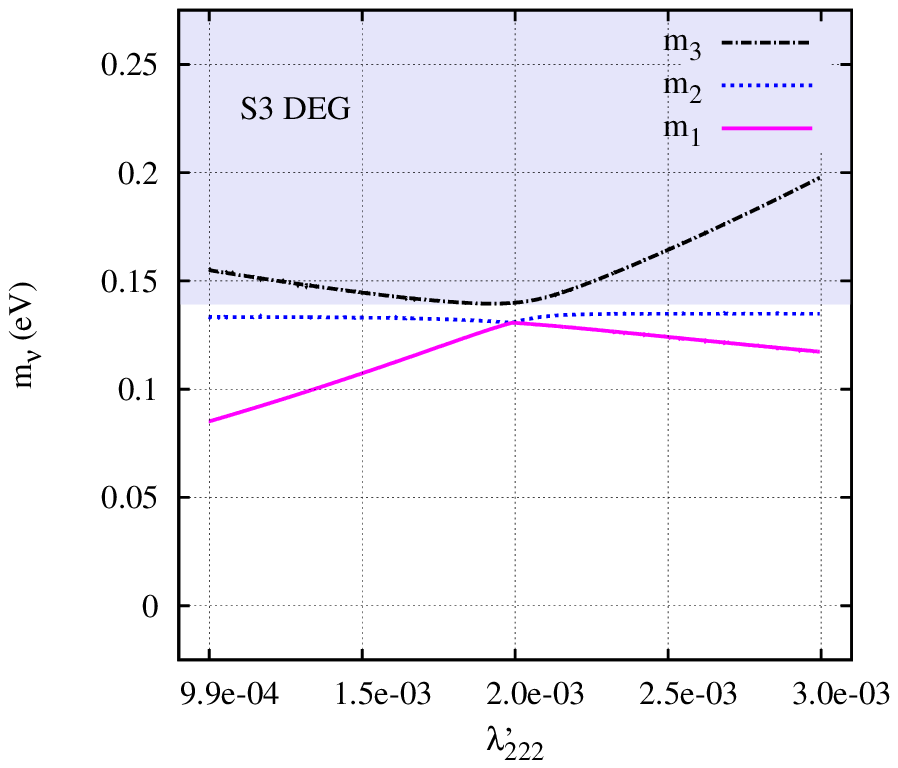,width=7.4cm}
  \epsfig{figure=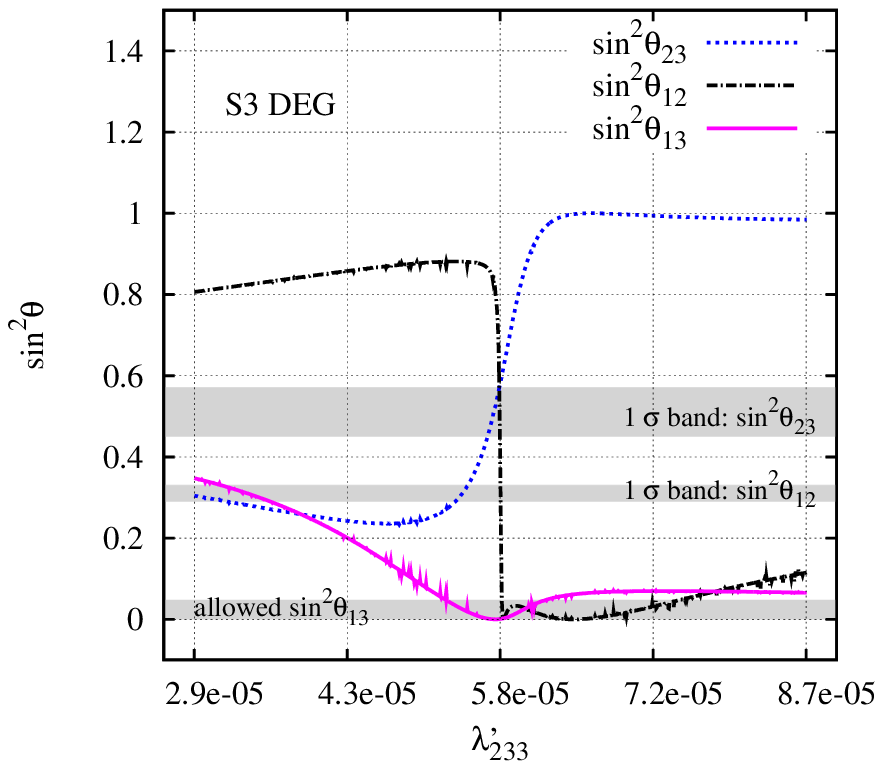,width=7.4cm}
  \epsfig{figure=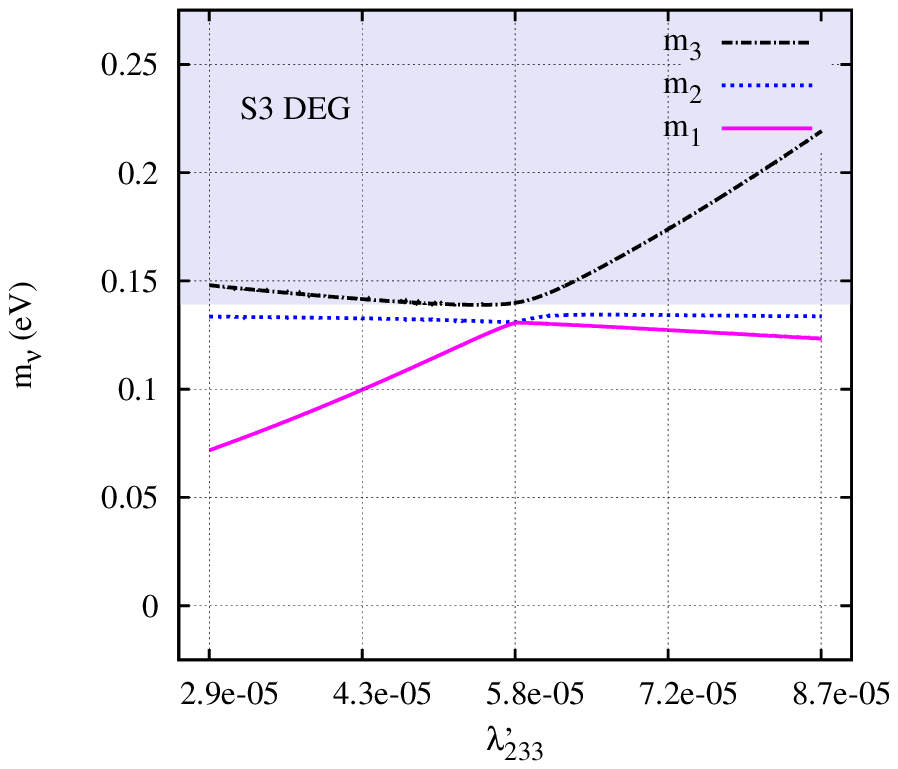,width=7.4cm}
  \caption{Variation of the mixing angles (left) and
  the mass eigenvalues (right) as functions of $\lamp_{222}$
  (top) and $\lamp_{233}$ (bottom) for scenario {\bf S3} DEG.  The
  best fit values for $\lamp_{222}$ and $\lamp_{233}$ are
  located at the centre of the plots.  On the plots of mixing angles,
  the grey bands are experimentally viable regions for (from top to
  bottom) ${\rm sin}^2\theta_{23}$, ${\rm sin}^2\theta_{12}$ and ${\rm
  sin}^2\theta_{13}$.  On the plots of mass eigenvalues, values inside
  the grey bands are disfavoured by cosmological considerations.
  \label{fig:SCEN3DEG}} \end{center}
\end{figure*}

To see how the experimental observables change as the LNV couplings
are varied, we show in Figs.\,\ref{fig:SCEN2NH}, \ref{fig:SCEN3IH} and
\ref{fig:SCEN3DEG} the variation of the mixing angles and masses 
as functions of $\lam'_{222}$.  Recall that the $\chi^2$ variation
of the fit for $\lam'_{222}$ is displayed in
Fig.~\ref{fig:chi2plots}.  For illustrative purposes these
figures also show the variation of another LNV coupling for each of
these scenarios, such that two sets of couplings, each corresponding
to one $M_{\alpha}$, are presented~\cite{ref:footnote13}.

We first discuss the scenario {\bf S2} NH, which is illustrated in
Fig.\,\ref{fig:SCEN2NH}.  In the upper two plots, one sees that the
variation of $\lamp_{222}$ mainly affects $\theta_{12}$ and somewhat
also $m_2$, whereas $\theta_{23}$ and $m_3$ are left relatively
unchanged. In the lower two plots, where $\lam_{211}$ is varied, the
observables are reversely affected.  This is because the two
non--zero mass matrices, $M_2 \sim m_{\nu 2}$ and $M_3\sim m_{\nu 3}$,
are controlled by the $\lam$ and $\lamp$ couplings separately
(\textit{i.e.} by the tree--level and loop level contribution,
respectively).  Obviously, in NH $\sin^2\theta_{12}$ is determined
only by $M_2$, whereas in IH and DEG, the form of $M_1$ is also
relevant. Therefore, NH is the easiest scenario to fit, because the
observables can be directly related to independent sets of couplings.
The mixing ${\rm sin}^2\theta_{13}$ remains practically unchanged due
to our ans\"atz in Eq.~(\ref{eq:ansatz}), which is designed to give a
tiny $\theta_{13}$.

For scenario {\bf S3} IH (Fig.\,\ref{fig:SCEN3IH}), we see that here,
no clean correlation exists between which LNV parameter is varied and
which observable is affected. $\theta_{12}$ and $m_2$ change
drastically and are affected by both $\lamp_{i22} \sim M_1$ and
$\lamp_{i33} \sim M_2$.  The sharp change in ${\rm sin}^2\theta_{12}$
around the best--fit point corresponds to ``cross--overs'' of mass
eigenstates $m_1$ and $m_2$ as $\lam'_{222}$ or $\lam'_{233}$ is
varied.  The fact that the best--fit solution lies in this steeply
changing region simply reflects the fact that for IH the two heavy
neutrinos have similar masses.  Incidentally, the small ``suppression''
at $\lamp_{222}\sim -8.4\cdot 10^{-4}$ in the corresponding $\chi^2$
plot in Fig.\,\ref{fig:chi2plots} near the best--fit point corresponds
to a region where $\Delta m^2_{21}$ coincides with the experimental
value during this cross--over.  However to a reasonable approximation
the flavour content of the two mass eigenstates are now swapped, hence
${\rm sin}^2\theta_{12}$ is different from its best--fit value.

On the other hand, it is clear that $m_3$ does not sit close to the
cross--over region.  Moreover, since $m_3$ basically contains only
$\mu$ and $\tau$ flavours around the best--fit region, the proportion
of $\mu$ and $\tau$ content of the other two mass eigenstates must be
the same in order for them to be orthogonal to $m_3$.  As a
consequence, the cross--over of these two states only changes
$\sin^2\theta_{23}$ mildly.  As in the case of {\bf S2} NH,
$\sin^2\theta_{13}$ is designed to have a tiny value.

For the scenario {\bf S3} DEG (Fig.\,\ref{fig:SCEN3DEG}), the fact
that the three mass scales are very close to each other means that
complete separation of the three contributions is in practice very
difficult.  As in {\bf S3} IH, the best--fit point lies close to a
region where cross--over of mass eigenstates take place.  In this
case, two cross--overs take place near the best--fit point.  For
example, the non--trivial variation of ${\rm sin}^2\theta_{12}$ with
$\lam'_{233}$ immediately to the right of the best--fit point
corresponds to a second cross--over of the mass eigenvectors.  The
fact that all three masses are quasi--degenerate also explains the
large transition of all three mixing angles.  In particular, even
though the coupling set is chosen to have a small ${\rm sin}^2\theta_
{13}$, immediately away from the best--fit point the mass ordering
is changed, resulting in the different ${\rm sin}
^2\theta_{13}$ behaviour compared with the NH and IH cases.

Furthermore, due to the strong fine--tuning, the $\chi^2$ suppression
expected as in the IH scenarios is buried within the rapidly
increasing $\chi^2$ value.  We note in passing that due to this
fine--tuning, the numerical results are less stable than those in the
NH and IH scenarios.  This results in the fluctuations seen in the
figures \cite{ref:footnote11}.

We now go on to discuss the scenarios {\bf S4}, which represent
scenarios with the smallest possible LNV couplings to still describe
the oscillation data correctly.  In the {\bf S4} NH scenario, recall
that the antisymmetry of the $\lam_{i33}$ couplings generates zeros in
$M_3$ which do not correspond to the ``texture zeros'' given in
Eq.~(\ref{eq:generalsplit}).  Therefore, linear combinations between
the different contributions to the neutrino masses (\textit{i.e.} 
between $M_2 \sim m_{\nu 2}$ and $M_3 \sim m_{\nu 3}$) are necessary
to obtain the desired oscillation parameters.  As a result, the ratio
of the couplings are not approximated by those displayed in
Eq.~(\ref{eq:simple_sol}) but instead by a linear combination of
these, \textit{cf.} Ref.~\cite{ref:footnote6}.  Still, the
behaviour of the observables when the relevant LNV couplings are
varied is similar to the scenarios discussed above.

In the {\bf S4} IH scenario, the $\lamp_{i33}$ couplings still roughly
follow the expected structure and magnitude as before in {\bf S1} to
{\bf S3} IH. However, the deviations are slightly larger because of
the presence of $\lam$ couplings.  In contrast to other IH scenarios,
in {\bf S4} IH, $M_{2}$ is generated at tree--level from $\lam_{i33}$
and $\lam_{322}$ instead of at 1--loop level from $\lamp_{i22}$.  The
absence of $\lam_{333}$, due to anti--symmetry of the first two
generation indices, means that $\lam_{322}$ (or $\lam_{311}$) is
needed to ``fill up'' the third row/column of the tree--level matrix
$M_2$. In this scenario, all diagonal third generation couplings are
used.  Consequently, the magnitude of our coupling set is the smallest
possible among the diagonal inverted hierarchy scenarios.

The ratio of the three $\lam$ couplings is approximately
\beq
(\lam_{133} : \lam_{233} : \lam_{322}) 
\;\sim\; (1 \,: \,1\, :\, -16) \,,\label{eq:ratios}
\eeq
which is expected as these couplings scales as $1/(Y_E)_{ii}$ ($i=2,3$).

We conclude in both the NH and the IH case that it is not possible to
push all LNV couplings below $\ord(10^{-5})$. However, at this order
of magnitude, displaced vertices might be observed at colliders,
depending on the benchmark point, \textit{cf.} Sect.~\ref{sec:Collider}.

\subsection{Off--diagonal LNV Scenarios}\label{sec:offdiagDiscussion}
In {\bf S5} we present the solutions for the two off--diagonal LNV
scenarios. We see that the NH off--diagonal solution, being an example
of non--hierarchical off--diagonal couplings, is very similar to the
diagonal NH solutions in structure, \textit{cf.}
Eq.~(\ref{eq:ratios}). Obviously, because here the generation
indices of the couplings are $i23/i32$ instead of $i22$ ({\bf
S2}) or $i33$ ({\bf S3}). The order of magnitude of the couplings is
somewhere between the solutions {\bf S2} and {\bf
S3}, mirroring the mass hierarchy in the down--quark sector.

In scenario {\bf S5} DEG, the $\lam_{231}$ coupling is much larger
than the other couplings, representing an example of a strongly
hierarchical off--diagonal scenario.  In fact,when performing the
\softsusy\, pre--iteration for our benchmark point, we found
$\lam_{231}$ to be of $\ord(1)$, which is inconsistent with the
requirement of perturbativity, and also violates the
low--energy bounds.

To reduce the size of this coupling, a different cSSM benchmark point
is therefore chosen.  Employing a larger $\tanb$ and also $\sgnmu=-1$
is useful, as the former implies larger down--type quark Yukawa
couplings, while the latter also increases certain loop contributions
to neutrino masses.  Of course, assuming a heavier mass spectrum is
also helpful.  In fact, a scan over the cSSM parameter space with the
condition $\lam_{231}\lesssim \ord(0.1)$, leads to the following
benchmark point (BP2): 
\bea 
\mhalf &=& 760 \textrm{ GeV}\,, \nonumber\\ 
\mo &=& 430 \textrm{ GeV}\,, \nonumber \\ 
\tanb &=& 40\,, \nonumber\\ 
\sgnmu &=& -1\,.  
\eea 

The $A_0^{(\lamp)}$ corresponding to this is 1059.2 GeV. The resulting
mass spectrum is displayed in Table~\ref{tab:BP2spectrum}.  Compare
with the original benchmark point BP, the sparticles in BP2 are
somewhat heavier than those in BP.  Also, while the LSP in BP is a
stau, the relatively small differences between $\mhalf$ and $\mo$ in
BP2 results in a neutralino LSP ($\tilde \chi^0_1$) instead. This
leads to distinctly different collider phenomenology, which will be
briefly discussed in the next section.

\begin{table}[t]
\begin{center}
\begin{tabular}{|c|cccc|}
\hline
Particles & \multicolumn{4}{|c|}{Masses (GeV)}\\
\hline
$\tilde{g}$& 1696 & & &\\
$\tilde{\chi}^{\pm}_{1},\tilde{\chi}^{\pm}_{2}$&599 & 798 & &\\
$\tilde{\chi}^{0}_{1},\tilde{\chi}^{0}_{2},\tilde{\chi}^{0}_{3},\tilde{\chi}^{0}_{4}$&\,\,320\,\,&\,\,599\,\,&\,\,785\,\,&\,\,799\,\,\\
$\tilde{u}_{1},\tilde{c}_{1},\tilde{t}_{1}$ &1593 &1593 &1431 &\\ 
$\tilde{u}_{2},\tilde{c}_{2},\tilde{t}_{2}$ &1536 &1535 &1281 &\\ 
$\tilde{d}_{1},\tilde{s}_{1},\tilde{b}_{1}$ &1595 &1595 &1427 &\\ 
$\tilde{d}_{2},\tilde{s}_{2},\tilde{b}_{2}$ &1530 &1530 &1358 &\\ 
$\tilde{e}_{1},\tilde{\mu}_{1},\tilde{\tau}_{1}$ &665 &665(663) &631(629) &\\ 
$\tilde{e}_{2},\tilde{\mu}_{2},\tilde{\tau}_{2}$ &516(510) &515 &382 &\\
$\tilde{\nu}_{e},\tilde{\nu}_{\mu},\tilde{\nu}_{\tau}$ &659 &659(657) &616(614) &\\
$h^0,A^0,H^0,H^{\pm}$&116&579&577&585\\
\hline
\end{tabular}
\caption{Mass spectrum of the benchmark point BP2 in the $\rp$
    conserving limit.  The notation is the same as
    Table~\ref{tab:BPspectrum}.  The values in brackets denote changes
    when the non--zero LNV couplings in {\bf S5} DEG is included.  As
    expected, the dominant coupling $\lam_{231}$ changes the second
    and third generation slepton and the (right--handed) selectron
    masses, but only by at most 1\%.
\label{tab:BP2spectrum}}
\end{center}
\end{table}

\subsection{Effects of changing the benchmark point} \label{sec:ChangesInBP}

So far, we have only considered scenarios under the assumption of
up--mixing in the quark sector and using the $A_0^{(\lamp)}$
minimum. In the rest of this section we briefly discuss changes which
occur when down--mixing is assumed or using the $A_0^{(\lam)}$ minimum
instead.
\bit
\item \textbf{$A_0^{(\lam)}$ minimum: } We consider as an example the
  scenario {\bf S2} NH. The best--fit LNV couplings for
  $A_0^{(\lamp)}=912.3$ GeV are given in the second row, first column
  of Table ~\ref{tab:results}. When using the $A_0^{(\lam)}$ minimum
  instead (given by $A_0^{(\lam)}=200.6$ GeV), the $\lamp_{i22}$
  couplings generate $M_2$ at tree--level whereas $M_3$ is generated
  by $\lam_{i11}$ at one--loop level (for the $A_0^{(\lamp)}$ it was
  the other way round). We obtain as a best fit
\barr
\lamp_{122} &=&  \:\:\:\,1.11 \cdot 10^{-5}\nonumber \\
\lamp_{222} &=&  \:\:\:\,1.49 \cdot 10^{-5}\nonumber \\
\lamp_{322} &= & -8.99 \cdot 10^{-6}\nonumber \\	 
\lam_{211} &=&  \:\:\:\,\vtext{\vmtext{1.53 \cdot 10^{-1}} \: [b3],[b5]}\nonumber \\
\lam_{311} &=&  \:\:\:\,\vtext{\vmtext{1.59 \cdot 10^{-1}} \: [b4]}
\earr
The decrease (increase) by a factor 10 of the $\lamp_{i22}$
($\lam_{i11}$) couplings reflects the typical hierarchy between the
tree--level and the one--loop neutrino mass of $\ord(10^2)$,
\textit{cf.}  Fig.~\ref{fig:A0plot}. In contrast to the original {\bf
S2} NH scenario, this scenario is not compatible with several
low--energy bounds as listed in Sect.~\ref{sec:lowEnergyBounds} due to
the larger $\lam_{i11}$ couplings.
\item \textbf{down--mixing: } When changing the quark mixing
  assumption from up--type to down--type mixing, \textit{cf.}
  Sec.~\ref{sec:FlavourBasis}, the LNV parameters are affected via RG
  running. However, the changes when running from the unification
  scale down to the electroweak scale are less than 1 percent for
  diagonal LNV couplings when switching from up--type to down--type
  mixing.  This is because for $\lam'$ couplings involving light
  generations (\textit{e.g.} $\lam'_{i11}$), RG running is dominated
  by gauge contributions.  For couplings involving the third
  generation (\textit{e.g.} $\lam'_{i33}$), the fact that the only
  significant mixing in the CKM matrix is between the first two
  generations implies that the effect of changing the quark mixing is
  also small.  The bilinear LNV couplings responsible for the tree
  level neutrino mass matrix are dynamically generated by $\lam$
  couplings, which are of course not affected directly by changes in
  the quark mixing assumptions.  In models where bilinear couplings
  are generated by $\lam'$ couplings, the effect of changing the quark
  mixing assumption is more complicated.
  
Note also that for non--diagonal couplings, the changes are expected
to be much larger than for diagonal couplings. This is because $Y_D$
is diagonal when assuming up--quark mixing, while non--zero
off--diagonal entries are present when down--quark mixing is assumed
instead.  We note that similar observations are made in
Ref.~\cite{Dreiner:2010ye}, where a single non--zero LNV coupling is
used to saturate the cosmological bound.

Nevertheless, these small changes for diagonal LNV couplings can still
be important, particularly for the IH and DEG scenarios, which are
sensitive to the exact values of the LNV parameters.  On top of that,
1--loop contributions involving light quark mass insertions can depend
sensitively on the quark mixing assumption.  For example, $(Y_D)_{11}$
changes by a factor of $\sim2$ when the mixing is changed, which
implies large changes in the loop contributions involving
$\lam'_{i11}$, which in turn will affect all mass ordering scenarios.
In contrast, $(Y_D)_{22}$ changes by a couple of percent, so the
impact through the mass insertion is relatively mild.
  
In principle, changing the mixing assumption, but retaining the same
coupling values, can affect $\chi^2$ dramatically, if the width $w$ of
the scenario is small.  As a numerical example consider a comparison
of the three scenarios depicted in Fig.~\ref{fig:chi2plots}. {\bf S2}
NH, involves $\lam'_{i22}$ with a width $w$ of $\ord(10\%)$.  Here
$\chi^2$ increases from $\sim 0$ in the up--mixing case to about $3$
in the down--mixing case.  In contrast, in {\bf S3} IH (DEG), where
the width is narrower than $1\%$ ($0.1\%$), changing the quark mixing
assumption leads to a $\chi^2$ change of 4 (more than 6) orders of
magnitude.  These changes can be compensated by refitting the LNV
couplings.  It is not surprising that refitting a subset of couplings
is sufficient.  For example, a refit of {\bf S3} IH yields:
\barr \label{eq:S3IH-downmix}
\lamp_{122} &=&  \:\:\:\,1.70 \cdot 10^{-3}\nonumber \\
\lamp_{222} &=& -8.80 \cdot 10^{-4}\nonumber \\
\lamp_{322} &= &  \:\:\:\,9.71 \cdot 10^{-4}\nonumber \\	 
\lamp_{133} &=&  \:\:\:\,3.11 \cdot 10^{-5}\nonumber \\
\lamp_{233} &=&  \:\:\:\,3.22 \cdot 10^{-5}\nonumber \\
\lamp_{333} &= & -3.32 \cdot 10^{-5} \,,
\earr
where the three $\lam'_{i22}$ are refitted.  A different solution with
a small $\chi^2$ can also be obtained by refitting $\lam'_{i33}$
alone.  The solution in Eq.\,(\ref{eq:S3IH-downmix}) differs from
the original up--type mixing solution by $\ord(10 
\%)$. This is what one might expect, bearing in mind that the changes
occurring in the CKM matrix from up--type to down--type mixing are
$\sim 20 \%$.

\eit

\section{Collider signatures}\label{sec:Collider}

The neutrino models we have found in the previous sections lead to
observable collider signatures.  Here, we shortly discuss
phenomenological implications at the LHC. Resonant slepton production
typically requires a coupling strength $\lam'_{i1 1}\gsim 10^{-3}$ for
incoming first generation quarks \cite{Dreiner:2000vf}. For higher
generation quarks an even larger coupling is required to compensate
the reduced parton luminosity.  In Table~\ref{tab:results}, we see
that our models do not satisfy this requirement. However, by
considering a scenario which combines aspects of {\bf S1} NH and {\bf S4} NH, it is possible to have a large
$\lam'_{211}$ while evading the low energy constraints, see \cite{ref:footnote18}.

Thus in most neutrino mass scenarios, squark and gluino production are
the dominant production mechanisms for supersymmetric particles at the
LHC, as in the $R_p$ conserving MSSM. Once produced, the squarks and
gluinos cascade decay in the detector to the LSP, via gauge
couplings. The final LHC signature is then determined by the exact
nature of the LSP and the LNV operators leading to the LSP decay.

For our benchmark point BP, we have a stau LSP $\tilde\tau_2$
\cite{ref:footnote17} and the NLSP is the lightest neutralino $\tilde
\chi_1^0$ with $m_{\tilde\tau_ 2}=163$ GeV and $m_{\tilde\chi_1^0}=
204$ GeV. A typical production process for our BP parameters is then 
given by
\beq
pp\rightarrow\tilde q\tilde q\rightarrow
qq\tilde\chi_1^0\tilde\chi_1^0\rightarrow
qq\tau\tau\tilde\tau_2\tilde\tau_2.
\label{eq:cascade_decay_stau_lsp}
\eeq
Here we have employed BR($\tilde \chi_1^0\rightarrow \tilde
\tau_2\tau)=1$, which is by far the dominant decay mode in our BP.
The LSP stau can normally decay via two-- and four--body modes
\cite{Dreiner:2008rv}. However, in our $B_3$ cSSM neutrino models
we always have a non--zero LNV operator which directly couples to the
stau LSP. Thus the stau will dominantly decay into two SM fermions and
the four--body decays of the stau LSP are highly suppressed. The
collider signatures can then be classified by the possible two--body
stau decay modes, as well as the stau decay length.  A recent detailed
discussion of stau LSP phenomenology at the LHC is given in
Ref.~\cite{Desch:2010gi}.  However, this focuses on four--body stau
decay modes.

For {\bf S2} NH, {\bf S3} NH, {\bf S5} NH and {\bf S3} DEG, we find
that $\lam_{311}$ is the dominant LNV operator which is relevant for
the tree--level two--body stau decay. Assuming the cascade decay in
Eq.~(\ref{eq:cascade_decay_stau_lsp}) we expect as the final state
collider signature
\beq
2j+2\ell+2\tau+2\nu.
\label{eq:lambdadecay}
\eeq
In this case $\ell=e$. Note that the final state charged leptons can
have the same electric charge, since the intermediate NLSP
neutralinos in Eq~(\ref{eq:cascade_decay_stau_lsp}) are Majorana
fermions. Like--sign dilepton signatures at the LHC in the context of
$\rpv$ have been studied extensively in the literature, see for example 
\cite{Dreiner:1993ba,likesign,Dreiner:2000vf}. Here we could in addition 
also make extra use of the final state tau leptons, as in
Ref.~\cite{Desch:2010gi}.

In {\bf S4} NH and {\bf S4} IH the stau LSP cannot decay via $\lamp_
{333}$, because it is kinematically forbidden, as $m_{\tilde\tau_2}
<m_{\rm top}$. Instead it will decay via $\lam_{133},\,\lam_{233}$,
or $\lam_{322}$ to a two--body leptonic final state. Hence, in both
scenarios the stau LSP decays into two leptons and we expect the same
signature as in Eq.~(\ref{eq:lambdadecay}). However, the couplings
have typical values of the order of $10^{-6}$--$10^{-5}$. The stau
lifetime is given by
\begin{eqnarray}
\tau_{\tilde\tau}&=&\left[
\Gamma(\tilde\tau\rightarrow f_1+f_2)
\right]^{-1}
=\frac{16\pi}{N_c\Lambda^2 m_{\tilde\tau_2}} \nonumber\\
&=& 3.3\cdot 10^{-15} \,\mathrm{sec}\frac{1}{N_c} 
\left(\frac{100\,\mathrm{GeV}}{m_{\tilde\tau_2}}\right)
 \left(\frac{10^{-5}}{\Lambda}\right)^2 \!\!.
\end{eqnarray}
Here $N_c$ is the colour factor. It is $3$ for $\lam'$ couplings and
$1$ for $\lam$ couplings. We have ignored any factors due to stau
mixing and have only considered one dominant decay mode
\cite{ref:footnote16}. The decay length is then given by
\begin{eqnarray}
L_{\tilde\tau_2}&=&\gamma\beta c\tau_{\tilde\tau_2} \nonumber\\
&=&\gamma\beta \cdot 10^{-6} m \cdot \frac{1}{N_c}\left(\frac{100\,\mathrm{GeV}}
{m_{\tilde\tau_2}}\right) \left(\frac{10^{-5}}{\Lambda}\right)^2.
\end{eqnarray}
In {\bf S4} NH the stau mass is 163 GeV and $c\tau_{\tilde\tau_2}\sim
3\mu$m.  The benchmark point BP implies that at the 14TeV LHC
$\gamma\beta$ is typically of $\ord({\rm few})$.  Therefore a small
fraction of events, with $\gamma\beta$ for one of the stau LSPs
near 10, could lead to detached vertices that are observable at the
LHC \cite{displaced}.

{\bf S3} IH is special. Here we just allow for non--zero $\lamp_{ijk}$
couplings. Hence, the stau LSP has only one hadronic two--body decay
mode via $\lamp_{322}$, $\tilde\tau_2\rightarrow c+s$. The final state
collider signature is
\beq
6j+2\tau.
\eeq
This is very difficult to observe. One must then consider other
cascades with intermediate first or second generation sleptons. These
lead to additional leptons in the final state. However, the
corresponding overall branching ratios are smaller.

For our benchmark point BP2, we have a neutralino LSP with $m_{\tilde
  \chi_1^0}=320$ GeV. 
A typical production process for BP2 is given by
\beq
pp\rightarrow\tilde q\tilde q\rightarrow
qq\tilde\chi_1^0\tilde\chi_1^0.
\label{eq:cascade_decay_chi_lsp}
\eeq
For {\bf S5} DEG, the dominant LNV coupling is $\lam_{231}$ and the
neutralino LSP decays via an off--shell slepton as
\begin{equation}
\tilde\chi_1^0\rightarrow \mu^\pm e^\mp\nu_\tau,\;\tau^\pm e^\mp\nu_\mu\,,
\end{equation}
and we did not distinguish between neutrinos and anti--neutrinos here.
We then expect the following event topologies
\beq
2j+2\nu+\left\{\begin{array}{l}
2\ell+2\tau,\\ 3\ell+1\tau,\\ 4\ell,
\end{array}
\right.
\eeq
where the branching ratios for all channels are roughly the same.

\section{Summary and Outlook}

Experimentally it is now well established that the neutrinos are
massive and have non--vanishing mixing angles. This requires physics
beyond the Standard Model. In this paper we have reanalyzed the
neutrino mass and mixing data in the light of supersymmetric R--parity
violating models. These automatically include lepton number violation,
and thus Majorana neutrino masses. One neutrino mass is generated at
tree--level via mixing with the conventional neutralinos. Any further
neutrino masses must arise at the one--loop level. We have improved
the accuracy of the neutrino mass and mixing angle computation, in
particular we have performed a full one loop calculation for the
sneutrino vacuum expectation values, on top of the one loop
corrections to the neutral fermion masses.  This computation is
implemented as an extension to the mass spectrum calculational tool
\softsusy~\cite{Allanach:2001kg,Allanach:2009bv}.

Most importantly, we have implemented also for the first time in the
construction of neutrino mass models, a mechanism to suppress the
tree--level masses compared to the corresponding 1--loop contribution.
This requires a tuning, but not fine--tuning, of the tri--linear soft
breaking $A_0$ parameter. This allows much larger flexibility in the
fitting procedure. It also allows for solutions with larger
lepton number violating couplings.

In this region of the $A_0$ parameter space, there are a large number
of possibilities to obtain the observed neutrino masses and mixings.
We have split our analysis into normal hierarchy (NH), inverted
hierarchy (IH) and degenerate (DEG) models. Furthermore we have mostly
focused on one benchmark point to fix the other cSSM parameters. We
have implemented all the relevant low--energy bounds on the
lepton number violating R--parity violating couplings. It
turns out these kill a significant number of the best--fit solutions
we find.

We have then considered five different scenarios, labelled \textbf{S1}
through \textbf{S5}. Scenarios \textbf{S1} through \textbf{S3} employ
diagonal lepton number violating couplings $\Lambda_{ijk},\;j=k$ and
the couplings are chosen to closely follow the structure of the
tri--bi maximal mixing solutions. The three scenarios correspond to
the three different possible generations $j=k=1,2,3$. \textit{Higher}
generations lead to \textit{smaller} lepton number violating
couplings, because the corresponding Higgs Yukawa couplings which also
enter the formulae are larger.

In looking for solutions, we then fit a small number of lepton number
violating couplings to the neutrino data. We need five couplings in
the NH case, six in the IH case and eight couplings for the degenerate
case. Our results are presented in Table~\ref{tab:results}.

Solutions with large couplings, $\Lambda=\mathcal{O}(10^{-2})$, are
mostly excluded by the low--energy bounds. In particular this kills
all \textbf{S1} models, as well as the IH and DEG models in the 
\textbf{S2} scenarios. The NH \textbf{S2}, as well as the NH and DEG
\textbf{S3} scenarios include $LL\bar E$ couplings of order $10^{-2}$.
All other remaining scenarios have couplings $10^{-3}$ or smaller.

Possible alternatives to the scenarios {\bf S1}, {\bf S2} and {\bf S3}
are presented in scenarios {\bf S4} and {\bf S5}.  The {\bf S4} models
assume ans\"atze with diagonal $\Lambda$ couplings but alternative
methods to obtain the neutrino masses, whereas the {\bf S5} models
employ off--diagonal $\Lambda$ couplings. We have not attempted to
construct IH \textbf{S5} nor DEG \textbf{S4} models.

All models lead to observable effects at colliders, as the LSP will
always decay in the detector. These have been discussed in detail
elsewhere. Characteristic of all neutrino models is that we should get
signatures which violate at least two lepton flavors. For the case of
\textbf{S4} scenarios, where all couplings are $|\Lambda|<5\cdot10^
       {-5}$ there could possibly be detached vertices.

In performing the numerical fit, we use a multistep procedure. We
start with initial values estimated from upper bounds on the neutrino
mass from WMAP. Then we perform separate pre--iterations for the
tree-- as well as for the 1--loop contributions in \softsusy. This
already gives a good estimate. The final solution is then found by
minimizing the $\chi^2$ function with the program package \minuit,
where all tree-- and 1--loop level contributions simultaneously
contribute to the neutrino mass matrix.  The degenerate scenarios
require some fine--tuning, thus we have implemented a Markov chain
Monte Carlo method to obtain the experimentally observed neutrino
data.

We find that all three neutrino mass hierarchies are possible, which
can contribute to $\nubb$ through the standard light neutrino
exchange.  However these simple models suggest normal hierarchy (NH)
might be preferred, so that the mass contribution to $\nubb$ will not
be probed by the next generation of $\nubb$ experiments.  All our
models involving $\lam'_{111}$ couplings strongly violate the limit
from its contribution to $\nubb$ through the so--called direct
neutralino/gluino exchange mechanism.  In other words, if $\nubb$ is
dominated by the direct exchange mechanism, $\lam'_{111}$ is unlikely
to contribute significantly to the neutrino masses.

Despite the tension between the neutrino mass contribution and the low
energy bounds, which favor large and small LNV couplings respectively,
$\lam$ couplings of $\ord(0.01)$ (e.g. {\bf S2, S3} NH) involving only
the first 2 lepton generations are allowed.  However, simultaneous
presence of (dominant) diagonal LNV couplings $\lam'_{i11}$ and
$\lam_{j11}$ appears to be difficult, at least with the assumed mass
spectrum BP.  Single coupling dominance, which many collider
studies usually assume, also appears to
be consistent with neutrino oscillation data  ({\bf S5} DEG).  It would therefore be
interesting to study collider implications of these models in more
detail in the future.

\begin{acknowledgments}
We thank Ben Allanach and Peter Wienemann for many useful discussions.
CHK and JSK would like to thank the Bethe Center of Theoretical
Physics and the Physikalisches Institut at the University of Bonn for
their hospitality. HKD and JSK would like to thank the SCIPP at the
University of California Santa Cruz for hospitality while part of this
work was completed.  CHK would also like to thank MPIK Heidelberg for
their hospitality.  This work has been supported in part by the Isaac
Newton Trust, the STFC, the Deutsche Telekom Stiftung, the
Bonn--Cologne Graduate School and the Initiative and Networking Fund
of the Helmholtz Association.
\newline 
\end{acknowledgments}


\begin{thebibliography}{99}
\bibitem{ref:footnote1} Note that at least two lepton flavors need to
  be violated in order to generate more than one neutrino
  mass. Therefore, one single LNV coupling will not be sufficient. For
  the discussion here, however, the simplifying picture of one LNV
  coupling is sufficient, since the arguments remain valid for more
  than one LNV coupling.
\bibitem{ref:footnote2} The scaling behaviour is non-trivial in
  general.  However, for the cases that we consider, the relevant
  penguin operators behave like dimension 6 operators.  We therefore
  use this approximation to compare with the (scaled) model limits
  whenever possible.
\bibitem{ref:footnote5} In the ``off-diagonal" scenarios, some
  deviation from this statement is necessary, as will be explained in
  Section~\ref{offDiagScen}.
\bibitem{ref:footnote7} Note that in principle, there would be 6
  possibilities. However, numerically the values of the LNV parameters
  are affected only at $\ord(1)$ level if we swap the assignment of $\lamp$ couplings to
  $c^{(1)}_i$ or $c^{(2)}_i$, \textit{i.e.}  $ c^{(1)}_i \sim
  \lamp_{i33}$, $c^{(2)}_i \sim \lamp_{i22}$ looks very similar to $
  c^{(1)}_i \sim \lamp_{i22}$, $ c^{(2)}_i \sim \lamp_{i33}$. This is obvious because
  the  $c^{(1)}_i$ and  $c^{(2)}_i$ differ from each other by maximally a factor 2.
\bibitem{ref:footnote8} Our choice to take the charged lepton mass
  matrix at the electroweak scale to be diagonal ensures that in good
  very approximation an off--diagonal coupling $\lam_{ijk}$ with $j \neq
  k$ does not generate a tree--level neutrino mass, since the bilinears
  $\kappa_i$ are generated proportionally to $\lam_{ijk} (Y_E)_{jk}$
  and are thus zero for $j\neq k$. This argument still roughly holds
  if there are small off--diagonal entries in the Higgs
  Yukawa coupling, so in approximation this  is also valid for couplings
  $\lamp_{ijk}$ with $j \neq k$, especially for the case of up--mixing.
\bibitem{ref:footnote15} Note that the coupling $\lam_{133}$
  contributes to both $M_2$ and $M_3$ due to the antisymmetry,
  $\lam_{133}\equiv -\lam_{313}$. We fix its value when fitting
  $M_2$. Therefore, effectively there are only 3 off-diagonal
  couplings to fit $M_3$, which is nonetheless sufficient. We set the
  $A_0$ minimum to $\lamp$, such that $M_2$ is generated at
  tree--level (leading to small $\lam_{i33}$ couplings) whereas $M_1$,
  $M_3$ are generated at loop level. For this reason, a strong
  hierarchy between $\lam_{313}$ and $\lam_{231}$ arises when fitting
  $(M_3)_{23} \sim\lam_{231} \lam_{313}$, because $\lam_{231}$ has to
  compensate for the smallness of $\lam_{313}$.  When now fitting
  $(M_3)_{22} \sim\lam_{231} \lam_{213}$, the large $\lam_{231}$
  coupling also leads to a strong hierarchy to $\lam_{213}$ in order
  to not exceed the experimental values (similarly for $(M_3)_{33}$
  and $\lam_{312}$).
\bibitem{ref:footnote10} This implies that one--loop corrections to
  the charged lepton mixing matrix would only indirectly influence the
  $\pmns$ matrix. Within the \softsusy\ iteration, the
  unification--scale Yukawa couplings are adjusted such that the
  charged lepton mixing matrix is always diagonal at the electroweak
  scale. One--loop corrections to $U_{\ell}$ would further alter
  (slightly) the unification-scale Yukawa couplings, which in turn
  affects the RGEs of the LNV parameters. However, these changes are
  negligible compared to the current experimental uncertainties in the
  neutrino sector, therefore we neglect one--loop corrections to the
  charged lepton mixing matrix.
\bibitem{ref:footnote9} For NH the couplings are smaller because the
  lighter neutrino mass is smaller in NH than in IH/DEG.
\bibitem{ref:footnote13} We refrain from showing an additional set for
the third $M_{\alpha}$ in the DEG case, because it does not give rise 
to any new insights.
\bibitem{ref:footnote11} In fact, the tolerance parameter in \softsusy
  needs to be set to high precision ($\ord(10^{-6})$) in order to
  produce results comparable among different platforms.
\bibitem{ref:footnote18} 
  For example, if we consider an "intermediate" scenario with
   $\tilde c^{(2)}_i \sim \lamp_{i11}$ and $\tilde c^{(3)}_i 
  \sim \lam_{i22}$, which can be achieved by
  using a linear combination of the original $c^{(\al)}_i$'s (similar
  to the construction of {\bf S4} NH), we can evade the bounds which
  exclude {\bf S1} NH and obtain a NH scenario with resonant smuon
  production. This is because this scenario leads to
  $\lamp_{211}=\mathcal{O}(10^{-3})$, whereas $\lamp_{111}\sim
  \ord(10^{-4})$ is sufficiently small in order to be consistent with
  {\bf [b8]}, due to the fact that $\tilde c^{(2)}_1/\tilde c^{(2)}_2
  \sim \ord(10^{-1})$.
\bibitem{ref:footnote17}
We employ here the convention used in \texttt{SOFTSUSY}, where
$\tilde\tau_1$ denotes the stau mass eigenstate which is primarily
left--handed and $\tilde\tau_2$ correspondingly right--handed.
\bibitem{ref:footnote16} 
For a primarily right--handed stau with a dominant $\lam_{ij3}$
  coupling, an extra factor of 0.5 should be included to account for the two
  final state configurations $\nu_il_j$ and $\nu_jl_i$.
\end{thebibliography}
\end{document}